\documentclass[12pt]{article}
\usepackage[paper=a4paper,margin=1in]{geometry}
\usepackage{amsmath}
\usepackage{amsfonts}
\usepackage{amssymb}
\usepackage{amsthm}
\usepackage{graphicx}
\usepackage{mathrsfs}  

\newcommand{\bi}{\bf i}

\numberwithin{equation}{section}

\begin{document}
\bibliographystyle{unsrt}

\title{Singularities in Einstein-conformally coupled Higgs cosmological 
models} 

\author{L\'aszl\'o B. Szabados,  Gy\"orgy Wolf\\
  Wigner Research Centre for Physics, \\
  H--1525 Budapest 114, P. O. Box 49, European Union}

\maketitle

\begin{abstract}
The dynamics of Einstein--conformally coupled Higgs field (EccH) system is 
investigated near the initial singularities in the presence of 
Friedman--Robertson--Walker symmetries. We solve the field equations 
asymptotically up to fourth order near the singularities analytically, and 
determine the solutions numerically as well. We found all the asymptotic, 
power series singular solutions, which are (1) solutions with a scalar 
polynomial curvature singularity but the Higgs field is \emph{bounded} 
(`Small Bang'), or (2) solutions with a Milne type singularity with bounded 
spacetime curvature and Higgs field, or (3) solutions with a scalar 
polynomial curvature singularity and diverging Higgs field (`Big Bang'). 
Thus, in the present EccH model there is a \emph{new kind} of physical 
spacetime singularity (`Small Bang'). We also show that, in a neighbourhood 
of the singularity in these solutions, the Higgs sector does not have any 
symmetry breaking instantaneous vacuum state, and hence then the 
Brout--Englert--Higgs mechanism does not work. The large scale behaviour of 
the solutions is investigated numerically as well. In particular, the 
numerical calculations indicate that there are singular solutions that 
cannot be approximated by power series.

\end{abstract}


\section{Introduction}
\label{sec:0}

The two most successful theories of the 20th century physics are General 
Relativity (see e.g. \cite{HE}) and the Standard Model of particle physics 
(see e.g. \cite{AL73}). In our previous paper \cite{Sz16} we investigated 
the origin of the rest masses of the \emph{classical} fields of the 
Einstein--Standard Model system in which the Higgs field is \emph{conformally 
coupled} to gravity (`Einstein--conformally coupled--Standard Model', or 
shortly, EccSM system). In this theory, in addition to the familiar Big Bang 
singularity, another (slightly less violent) singularity may also emerge 
(`Small Bang'). In the latter all the matter field variables are 
\emph{finite}, and this singularity corresponds to a special, finite value 
of the pointwise norm of the Higgs field. In \cite{Sz16} we primarily 
concentrated on how the Brout--Englert--Higgs (or shortly BEH) mechanism 
works in this system. We found that there could be extreme gravitational 
situations in which the system does not have any vacuum state, even 
instantaneous ones, and hence the notion of rest mass of the Higgs field 
cannot be introduced at all and the gauge and the fermion fields are still 
massless. When the system has vacuum states, then these states are only 
\emph{instantaneous} and \emph{necessarily gauge symmetry breaking}. Then, 
via the BEH mechanism, the fields get rest mass. Therefore, the rest mass 
(and electric charge in the Weinberg--Salam model) has a non-trivial 
genesis \emph{after} the initial singularity. 

To derive these results it was enough to consider only the 
\emph{kinematical} structure of the EccSM system (using only the 
\emph{constraint} equations) \cite{Sz16}, but we did not investigate its 
\emph{dynamics} (i.e. we did not use the \emph{evolution} equations). In the 
present paper our aim is two-fold: (1) to clarify the dynamics of the model 
near both the points where the Higgs field takes the critical value above 
(where the Small Bang is expected to be present) and the Big Bang singularity, 
and, in particular, to demonstrate that the Small Bang is not fictitious, but 
it is a genuine scalar polynomial curvature singularity; and (2) to justify 
the key observation of \cite{Sz16} that the rest masses of the classical 
fields could emerge in a non-trivial `phase transition' in the very early 
period of the history of the Universe in a dynamical process \emph{after} 
the initial singularity. In this very early era the dominant matter field 
is the Higgs field. Thus, for the sake of simplicity, we consider only the 
Einstein--conformally coupled Higgs (or shortly EccH) system in which the 
Higgs field is a single real self-interacting scalar field $\Phi$ in the 
presence of Friedman--Robertson--Walker (or FRW) symmetries. We determine 
all the asymptotic (power series) solutions near the singularities. Since 
the asymptotic, power series techniques are appropriate to determine the 
behaviour of the solutions only in the very small neighbourhoods of the 
singularities, to see the structure of the solutions on larger scales, we 
should find numerical solutions as well. 

We found all the asymptotic singular (power series) solutions of the field 
equations, which are (1) solutions with a scalar polynomial curvature 
singularity but in which the Higgs field is bounded (`Small Bang'), or (2) 
solutions with a Milne type singularity \cite{ES} with bounded spacetime 
curvature and Higgs field, or (3) solutions with a scalar polynomial 
curvature singularity and diverging Higgs field (`Big Bang'). The solutions 
with a Small Bang or a Big Bang singularity form a 1-parameter family of 
solutions for any value $k=\pm1$ of the discrete cosmological parameter. 
The solutions for $k=0$ are determined (up to an overall scale factor) by 
the parameters of the theory. The solutions with a Milne type singularity 
exist only for $k=-1$, but these depend on a continuous parameter. The 
latter solutions can be continued through the (fictitious) Milne type 
singularity, describing a contracting and then expanding universe. Already 
these asymptotic solutions answer the questions above: (1) the Small Bang 
singularity is a genuine physical spacetime singularity, and (2) in a 
neighbourhood of the initial singularity there is, indeed, a very early 
period in the history of the Universe when the BEH mechanism does 
\emph{not} work and hence the fields of the Standard Model could not get 
non-zero rest mass via the BEH mechanism. 

We solve the equations of motion numerically, too. We show that the 
approximate, power series solutions can be extended from $10^{-3}$ to $10^{22}$ 
Planck times, and this time interval has an overlap with the era governed by 
the weak interactions. Moreover, we found that, in addition to the power 
series type singular solutions, there are other singular solutions that 
cannot be approximated by any power series in a neighbourhood of the 
singularity.

In section \ref{sec:1} we recall briefly the key points of the EccH model 
with the FRW symmetries. Section \ref{sec:2} is devoted to the asymptotic 
(power series) solutions in which the Higgs field remain bounded; while the 
solutions with diverging Higgs field are determined in section \ref{sec:3}. 
The numerical results are presented and discussed in section \ref{sec:4}; 
and the results and the main messages of the paper are summarized in section 
\ref{sec:5}. 

Our conventions are those of \cite{Sz16}. In particular, the signature of 
the spacetime metric is $(+,-,-,-)$ and Einstein's equations take the form 
$R_{ab}-\frac{1}{2}Rg_{ab}=-\kappa T_{ab}-\Lambda g_{ab}$. Here $\Lambda$ is 
the cosmological constant and $\kappa:=8\pi G$ with Newton's gravitational 
constant $G$.


\section{The EccSM system with FRW symmetries}
\label{sec:1}

\subsection{The field equations}
\label{sub-1.1}

Let $\Sigma_t:=\{t={\rm const}\}$ be the foliation of the FRW symmetric 
spacetime by the transitivity surfaces of the isometries, where $t$ is the 
proper time coordinate along the integral curves of the future pointing unit 
normals of the hypersurfaces $\Sigma_t$ (see e.g. \cite{HE}). Thus the lapse 
is $N=1$. Let $S=S(t)$ be the (strictly positive) scale function for which 
the induced metric on $\Sigma_t$ is $h_{ab}=S^2{}_1h_{ab}$, where ${}_1h_{ab}$ 
is the standard negative definite metric on the unit 3-sphere, the flat 
3-space and the unit hyperboloidal 3-space, respectively, for $k=1,0,-1$. 
The extrinsic curvature of the hypersurfaces is $\chi_{ab}=(\dot S/S)h_{ab}$, 
where over-dot denotes derivative with respect to $t$, and hence its trace 
is $\chi=3\dot S/S$. The curvature scalar of the intrinsic Levi-Civita 
connection is ${\cal R}=6k/S^2$. In the initial value formulation of 
Einstein's theory the initial data are $h_{ab}$ and $\chi_{ab}$, and hence in 
the present case $S$ and $\dot S$, restricted by the constraint equations. 

For the metric with FRW symmetries Einstein's equations are well known 
\cite{HE} to reduce to 

\begin{equation}
3\bigl(\frac{\dot S}{S}\bigr)^2=\Lambda+\kappa\varepsilon-3\frac{k}{S^2}, 
\hskip 20pt
3\frac{\ddot S}{S}=\Lambda-\frac{1}{2}\kappa\bigl(\varepsilon+3P\bigr), 
\label{eq:1.1}
\end{equation}
where $\varepsilon$ is the energy density and $P$ is the isotropic pressure 
in the energy-momentum tensor of the matter fields. The first of these 
equations is the Hamiltonian constraint, while the second is the evolution 
equation. (The momentum constraint is satisfied identically.) 

If the fields of the matter sector of the EccSM system are required to be 
invariant under the isometries of the spacetime, then all the fields with 
spatial vector or spinor index must be vanishing and the Higgs field 
$\Phi^{\bi}$ and its canonical momentum, $\Pi^{\bi}=\dot\Phi^{\bi}+\frac{1}{3}
\chi\Phi^{\bi}$, must be constant on the hypersurfaces $\Sigma_t$. Thus, the 
EccSM system restricted by the FRW symmetries reduces to the 
Einstein--conformally coupled Higgs (EccH) system. For the sake of 
simplicity, instead of a non-trivial multiplet of scalar fields, we consider 
the Higgs field only to be a single real scalar field $\Phi$, the gauge 
group to be $\mathbb{Z}_2$ acting on the Higgs field as $\Phi\mapsto-\Phi$, 
and the Lagrangian for the Higgs field is 

\begin{equation*}
{\cal L}_H:=\frac{1}{2}g^{ab}(\nabla_a\Phi)(\nabla_b\Phi)-\frac{1}{12}R\Phi^2-
\frac{1}{2}\mu^2\Phi^2-\frac{1}{4}\lambda\Phi^4.
\end{equation*}
Here $R$ is the curvature scalar of the spacetime, $\lambda>0$ is the Higgs 
self-interaction and $\mu^2<0$ is the mass parameter. (In the $\hbar=c=1$ 
units the numerical value of the various constants of the model are $\Lambda
=10^{-58}cm^{-2}$, $6/\kappa=8.6\times10^{64}cm^{-2}$, $\lambda=1/8$ and $\mu^2=
-1.8\times10^{31} cm^{-2}$.) A simple calculation gives that the trace of the 
energy-momentum tensor of the Higgs field is $\mu^2\Phi^2$. Then, using the 
trace of Einstein's equation, $R=4\Lambda+\kappa\mu^2\Phi^2$, the field 
equation for the Higgs field takes the form 

\begin{equation}
\ddot\Phi+3\frac{\dot S}{S}\dot\Phi=-\bigl(\mu^2+\frac{2}{3}\Lambda\bigr)
\Phi-\bigl(\lambda+\frac{1}{6}\kappa\mu^2\bigr)\Phi^3.  \label{eq:1.2}
\end{equation}
The initial data for the evolution equations is the quadruplet $(\Phi,S;\dot
\Phi,\dot S)$, or, equivalently, $(\Phi,S;\Pi,\chi)$, subject to the 
constraint part of (\ref{eq:1.1}). Thus the configuration space ${\cal Q}$ 
of the dynamical system (\ref{eq:1.1})-(\ref{eq:1.2}) is the set of the pairs 
$(\Phi,S)$, where $S>0$; while its velocity and momentum phase spaces, $T
{\cal Q}$ and $T^*{\cal Q}$, are the sets of the quadruplets $(\Phi,S;\dot
\Phi,\dot S)$ and $(\Phi,S;\Pi,\chi)$, respectively, with $S>0$. The first of 
(\ref{eq:1.1}) yields the constraint hypersurface, $C(\Phi,S,\Pi,\chi)=0$, 
in $T^*{\cal Q}$. On time intervals in which $\dot\chi$ is non-zero, $\chi$ 
can also be used as a natural time variable (`York time'), and hence the 
hypersurfaces $\Sigma_t$ of the foliation can be labelled by $\chi$. 

\subsection{The energy density}
\label{sub-1.2}

Calculating the energy-momentum tensor from the matter action based on 
${\cal L}_H$ and using Einstein's equations, for the energy density of the 
Higgs field on the hypersurfaces $\Sigma_t$ we obtain 

\begin{equation}
\varepsilon=\frac{1}{2}\frac{1}{1-\frac{1}{6}\kappa\Phi^2}\Bigl(\Pi^2+
\bigl(\mu^2+\frac{1}{3}\Lambda-\frac{1}{9}\chi^2\bigr)\Phi^2+\frac{1}{2}
\lambda\Phi^4\Bigr), \label{eq:1.3}
\end{equation}
the momentum density is zero, and the spatial stress is pure trace, in which 
the isotropic pressure is $P=\frac{1}{3}\varepsilon-\frac{1}{3}\mu^2\Phi^2$ 
(see \cite{Sz16}). Hence $\varepsilon=\varepsilon(\Phi,\Pi,\chi)$, i.e. it 
does not depend on the gravitational configuration variable $S$. Using 
$3P=\varepsilon-\mu^2\Phi^2$, a piece of the `conservation law', $0=(\nabla
_aT^a{}_b)t^b$, takes the form $\frac{\rm d}{{\rm d}t}(\varepsilon S^4)=
\frac{1}{4}\mu^2\Phi^2\frac{\rm d}{{\rm d}t}(S^4)$. If $\mu^2$ were zero, 
then this equation would yield $\varepsilon(t)={\rm const}\,S^{-4}(t)$, which 
is the familiar time dependence of the energy density in the radiation filled 
standard cosmological models. 

In the momentum phase space the energy density has two singularities: The 
first is when $\Phi^2\rightarrow\infty$ or $\Pi^2\rightarrow\infty$ (Big 
Bang), and the other could be when $\Phi^2\rightarrow6/\kappa$. In fact, in 
the $\Phi=\pm\sqrt{6/\kappa}$, $S={\rm const}$ 2-planes of $T^*{\cal Q}$ the 
energy density is finite, viz. $\Pi^2/2-9\lambda/\kappa^2$, precisely only 
on the two hyperbolas $\chi^2-3\kappa\Pi^2/2=\chi^2_c$, where $\chi^2_c:=9(
\mu^2+\Lambda/3+3\lambda/\kappa)$. (In the $\hbar=c=1$ units $\frac{1}{9}
\chi^2_c\simeq5.4\times10^{63}cm^{-2}$.) Thus, apart from these lines, any 
point of the $\Phi^2=6/\kappa$, $S={\rm const}$ 2-planes is a singularity of 
$\varepsilon$. (As we will see, the Small Bang will be such a singularity.) 
For given $\chi$ the energy density (i.e. $\varepsilon$ as a function of 
$\Phi$ and $\Pi$) can have local minima only if $\chi^2<\chi^2_c$. These 
minima are at $\Pi=0$ and $\Phi=\Phi_v$ given by 

\begin{equation}
\Phi^2_v=\frac{6}{\kappa}\Biggl(1-\sqrt{1+\frac{\kappa}{3\lambda}\bigl(\mu^2
+\frac{1}{3}\Lambda-\frac{1}{9}\chi^2\bigr)}\Biggr)=\frac{6}{\kappa}\Biggl(1-
\sqrt{\frac{\kappa}{27\lambda}}\sqrt{\chi^2_c-\chi^2}\Biggr). \label{eq:1.4}
\end{equation}
Clearly, $\Phi^2_v\rightarrow6/\kappa$ if $\chi\rightarrow\chi_c$ and $\lim
_{\chi\rightarrow0}\Phi^2_v$ is also finite. A simple calculation shows that 
$({\rm d}\Phi_v/{\rm d}\chi)$ tends to $\infty$ if $\chi\rightarrow\chi_c$, 
and to zero if $\chi\rightarrow0$. Hence the graph of $\Phi_v(\chi)$ in the 
$(\Phi,\chi)$--plane of the phase space is confined to the square $\Phi^2
\leq6/\kappa$, $\chi^2<\chi^2_c$; and the states of minimal energy density 
are all on the 2-surfaces $(\Phi,S;\Pi,\chi)=(\Phi_v(\chi),S;0,\chi)$ in $T^*
{\cal Q}$ for $S>0$ and $\chi^2<\chi^2_c$. At $\chi=\chi_c$ the curve $\Phi_v
(\chi)$ reaches the $\Pi=0$ point of the hyperbola of non-singular points of 
the energy density on the $\Phi=\sqrt{6/\kappa}$, $S={\rm const}$ 2-plane. 
The minimal value of the energy density at the point $(\Phi_v,S;0,\chi)$ is 
$\varepsilon_v(\chi)=-\frac{1}{4}\lambda\Phi^4_v(\chi)$. (For a more 
detailed discussion, see \cite{Sz16}.) 

\subsection{The vacuum states}
\label{sub-1.3}

The usual notion of spacetime vacuum states of field theories cannot be 
introduced in the EccSM system: There are \emph{no solutions} of \emph{all} 
the field equations which would admit \emph{maximal} spacetime symmetry and 
minimize the energy \emph{density} at the same time \cite{Sz16}. Hence, these 
criteria in the definition of vacuum states should be relaxed, e.g. to 
\emph{instantaneous} states on the spacelike hypersurfaces $\Sigma_t$ 
satisfying only the \emph{constraint} (rather than all the field) equations 
and minimizing the energy \emph{functional}. In particular, in the presence 
of FRW symmetries, these states on the hypersurface labelled by $\chi$ are 
those $(\Phi_v,S_v;0,\chi)$ in which the Higgs field configurations $\Phi_v=
\Phi_v(\chi)$ are given by (\ref{eq:1.4}) and $S_v=S_v(\chi)$ solves the 
Hamiltonian constraint $\frac{1}{3}\chi^2=\Lambda-\lambda\Phi^4_v/4-3k/S^2_v$. 
The solution of the latter is $S^2_v=-k/(\lambda\Phi^2_v+\mu^2)$. However, 
since $S^2_v$ must be non-negative and $\lambda\Phi^2_v(\chi)+\mu^2>0$ holds 
for any $\chi\in(-\chi_c,\chi_c)$, it follows that $k=-1$, and hence 

\begin{equation}
S^2_v=\frac{1}{\mu^2+\lambda\Phi^2_v(\chi)}. \label{eq:1.5}
\end{equation}
This is finite and bounded on the whole interval $(-\chi_c,\chi_c)$. Therefore, 
the constraint equations (actually, the Hamiltonian constraint) can be solved 
\emph{globally} on $\Sigma_t$ (`global instantaneous vacuum states') precisely 
when the discrete parameter $k$ in the field equations (\ref{eq:1.1}) is 
$k=-1$. If $\chi^2\geq\chi^2_c$, then the energy density (\ref{eq:1.3}) is 
\emph{not} bounded from below, and hence no vacuum state (symmetric or 
symmetry breaking) exists. Therefore, the rest mass of the Higgs field is 
\emph{not defined at all}. (In this case, in the more general EccSM system 
the BEH mechanism does not work, and the gauge and fermion fields remain 
massless.) If $\chi^2<\chi^2_c$, then the vacuum states are symmetry breaking, 
the rest mass of the Higgs field is well defined, and, in the EccSM system, 
the gauge and spinor fields get rest masses via the BEH mechanism. The time 
dependence of $\Phi^2_v$ (via the time dependence of $\chi$) yields time 
dependence of the rest masses \cite{Sz16}. The 1-parameter family of these 
instantaneous vacuum states does \emph{not} solve the \emph{evolution} 
equations: The evolution equations take instantaneous vacuum states into 
non-vacuum states, and non-vacuum sates may develop into instantaneous 
vacuum states.


\section{Asymptotic solutions with bounded Higgs field}
\label{sec:2}

In this section we determine \emph{all} the asymptotic power series solutions 
of the field equations (\ref{eq:1.1})-(\ref{eq:1.2}) when the Higgs field is 
bounded. Since primarily we are interested in solutions that are singular at 
$t=0$, and since the energy density $\varepsilon$ is \emph{formally} singular 
at $\Phi^2=6/\kappa$, we should consider the disjoint cases when $\Phi^2(t)
\rightarrow 6/\kappa$, and when $\Phi^2(t)\not\rightarrow 6/\kappa$ in the 
$t\rightarrow0$ limit. Hence, in the former case, we should consider the 
possibilities when $S^2(t)\rightarrow0$ and $S^2(t)\rightarrow S_0>0$, but in 
the latter only when $S^2(t)\rightarrow0$. (In the second case the solution 
with $S^2(t)\rightarrow S_0>0$ would be \emph{a priori} regular at $t=0$.) 
Thus, we write the Higgs field $\Phi(t)$ and the scale function $S^2(t)$ as 

\begin{eqnarray}
\Phi&\!\!\!\!=&\!\!\!\!\sqrt{\frac{6}{\kappa}}\Bigl(\phi_0+\phi_1t+\phi_2t^2+
  \phi_3t^3+\phi_4t^4+{\cal O}(t^5)\Bigr), \label{eq:2.1a} \\
S^2&\!\!\!\!=&\!\!\!\!S_0+S_1t+S_2t^2+S_3t^3+S_4t^4+{\cal O}(t^5), 
  \label{eq:2.1b}
\end{eqnarray}
where $\phi_0,...,\phi_4$ and $S_0,...,S_4$ are real constants. The details 
of the analysis depend on the order of the first non-zero expansion 
coefficient, say $S_n$, in (\ref{eq:2.1b}). Thus we write the scale function 
as $S^2(t)=S_nt^n+S_{n+1}t^{n+1}+...$, where $n\geq0$. Moreover, if $n>0$, then 
we allow $n$ to be a positive real, rather than only a positive integer.

\subsection{The field equations}
\label{sub-2.1}

Substituting the expansion of $\Phi$ and $S^2$ above into the evolution 
equation (\ref{eq:1.2}) for the Higgs field, we obtain 

\begin{eqnarray}
0\!\!\!\!&=\!\!\!\!&\frac{3n}{2}\phi_1\frac{1}{t}+\Bigl((2+3n)\phi_2+
  \frac{3}{2}\phi_1\frac{S_{n+1}}{S_n}+(\mu^2+\frac{2}{3}\Lambda)\phi_0+(\mu^2
  +6\frac{\lambda}{\kappa})\phi^3_0\Bigr)+ \nonumber \\
\!\!\!\!&+\!\!\!\!&\Bigl((6+\frac{9}{2}n)\phi_3+3\phi_2\frac{S_{n+1}}{S_n}+3
  \phi_1\bigl(\frac{S_{n+2}}{S_n}-\frac{1}{2}\frac{S^2_{n+1}}{S^2_n}\bigr)+\phi_1
  \bigl(\mu^2+\frac{2}{3}\Lambda\bigr)+3\phi^2_0\phi_1\bigl(\mu^2+6
  \frac{\lambda}{\kappa}\bigr)\Bigr)t+ \nonumber \\
\!\!\!\!&+\!\!\!\!&\Bigl(6(n+2)\phi_4+\frac{9}{2}\phi_3\frac{S_{n+1}}{S_n}+6
  \phi_2\bigl(\frac{S_{n+2}}{S_n}-\frac{1}{2}\frac{S^2_{n+1}}{S^2_n}\bigr)+
  \frac{9}{2}\phi_1\frac{S_{n+3}}{S_n}-\frac{9}{2}\phi_1\frac{S_{n+1}S_{n+2}}
  {S^2_n}+ \nonumber \\
\!\!\!\!&{}\!\!\!\!&+\frac{3}{2}\phi_1\frac{S^3_{n+1}}{S^3_n}+(\mu^2+\frac{2}
  {3}\Lambda)\phi_2+3(\mu^2+6\frac{\lambda}{\kappa})(\phi_0\phi^2_1+\phi^2_0
  \phi_2)\Bigr)t^2+{\cal O}(t^3). \label{eq:2.2}
\end{eqnarray}
Similarly, the sum of the two Einstein equations in (\ref{eq:1.1}) is 

\begin{eqnarray}
0\!\!\!\!&=\!\!\!\!&n(n-1)\frac{1}{t^2}+2n\frac{S_{n+1}}{S_n}\frac{1}{t}+2
  \Bigl((1+2n)\frac{S_{n+2}}{S_n}-n\frac{S^2_{n+1}}{S^2_n}-\mu^2\phi^2_0-
  \frac{2}{3}\Lambda)\Bigr)+ \nonumber \\
\!\!\!\!&+\!\!\!\!&2\Bigl(3(n+1)\frac{S_{n+3}}{S_n}-(3n+1)\frac{S_{n+1}S_{n+2}}
  {S^2_n}+n\frac{S^3_{n+1}}{S^3_n}-2\mu^2\phi_0\phi_1\Bigr)t+{\cal O}(t^2)+ 
  \nonumber \\
\!\!\!\!&+\!\!\!\!&\frac{2k}{S_n}\frac{1}{t^n}-2k\frac{S_{n+1}}{S^2_n}\frac{1}
  {t^{n-1}}+2k\Bigl(\frac{S^2_{n+1}}{S^3_n}-\frac{S_{n+2}}{S^2_n}\Bigr)\frac{1}
  {t^{n-2}}-\nonumber \\
\!\!\!\!&-\!\!\!\!&2k\Bigl(\frac{S^3_{n+1}}{S^4_n}-2\frac{S_{n+1}S_{n+2}}{S^3_n}+
  \frac{S_{n+3}}{S^2_n}\Bigr)\frac{1}{t^{n-3}}+{\cal O}(t^{-n+4}), \label{eq:2.3}
\end{eqnarray}
while their difference is 

\begin{eqnarray}
0\!\!\!\!&=\!\!\!\!&\frac{4\kappa}{3}\varepsilon-\frac{n}{t^2}+\Bigl(2
  \frac{S_{n+2}}{S_n}-\frac{S^2_{n+1}}{S^2_n}-2\mu^2\phi^2_0\Bigr)+ \nonumber \\
\!\!\!\!&+\!\!\!\!&2\Bigl(3\frac{S_{n+3}}{S_n}-3\frac{S_{n+1}S_{n+2}}{S^2_n}+
  \frac{S^3_{n+1}}{S^3_n}-2\mu^2\phi_0\phi_1\Bigr)t+{\cal O}(t^2)-\frac{2k}{S_n}
  \frac{1}{t^n}+2k\frac{S_{n+1}}{S^2_n}\frac{1}{t^{n-1}}- \nonumber \\
\!\!\!\!&-\!\!\!\!&2k\Bigl(\frac{S^2_{n+1}}{S^3_n}-\frac{S_{n+2}}{S^2_n}\Bigr)
  \frac{1}{t^{n-2}}+2k\Bigl(\frac{S^3_{n+1}}{S^4_n}-2\frac{S_{n+1}S_{n+2}}{S^3_n}
  +\frac{S_{n+3}}{S^2_n}\Bigr)\frac{1}{t^{n-3}}+{\cal O}(t^{-n+4}). \label{eq:2.4}
\end{eqnarray}
The advantage of these combinations is that the energy density $\varepsilon$ 
appears only in the second. Thus (\ref{eq:2.2}) and (\ref{eq:2.3}) can be 
evaluated without the explicit form of $\varepsilon$. The actual structure 
of $\varepsilon$ depends on the order of the first non-trivial expansion 
coefficient in (\ref{eq:2.1a}). Thus we calculate its asymptotic expansion 
in the specific cases.

\subsection{The asymptotic solutions with a Small Bang  singularity}
\label{sub-2.2}

First let us consider the case $\Phi^2\rightarrow 6/\kappa$, and choose 
$\phi_0=1$ (rather than $\phi_0=-1$). Let us suppose that $S_0=0$, i.e. now 
we search for a solution in which the spacetime geometry could be singular. 
Let us start with the assumption $n=1$, i.e. $S_1\not=0$ in the expansion 
(\ref{eq:2.1b}). Then equations (\ref{eq:2.2})-(\ref{eq:2.3}) with $\phi_0
=1$ and $n=1$ yield 

\begin{eqnarray}
&{}&S_2=-k, \label{eq:2.5a} \\
&{}&S_3=\frac{1}{3}(\mu^2+\frac{2}{3}\Lambda)S_1, \label{eq:2.5b} \\
&{}&S_4=-\frac{1}{6}(\mu^2+\frac{2}{3}\Lambda)k, 
  \label{eq:2.5c} \\
&{}&\phi_1=0, \label{eq:2.5d} \\
&{}&\phi_2=-\frac{2}{5}(\mu^2+\frac{\Lambda}{3}+3\frac{\lambda}{\kappa}), 
  \label{eq:2.5e} \\
&{}&\phi_3=-\frac{4}{35}(\mu^2+\frac{\Lambda}{3}+3\frac{\lambda}{\kappa})
  \frac{k}{S_1}, \label{eq:2.5f} \\
&{}&\phi_4=\frac{2}{15}(\mu^2+\frac{\Lambda}{3}+3\frac{\lambda}{\kappa})
  \Bigl(\mu^2+\frac{\Lambda}{3}+3\frac{\lambda}{\kappa}-\frac{5}{7}\frac{k^2}
  {S^2_1}\Bigr); \label{eq:2.5g}
\end{eqnarray}
while the difference of the two Einstein equations is 

\begin{equation}
0=-\frac{4\kappa}{3}\varepsilon+\frac{1}{t^2}+\frac{2k}{S_1}\frac{1}{t}+\bigl(
\frac{4}{3}\mu^2-\frac{4}{9}\Lambda+3\frac{k^2}{S_1^2}\bigr)-\frac{k}{S_1}
\Bigl(\frac{4}{3}(\mu^2+\frac{2}{3}\Lambda)-4\frac{k^2}{S^2_1}\Bigr)t+{\cal O}
(t^2). \label{eq:2.6}
\end{equation}
Using (\ref{eq:2.1a})-(\ref{eq:2.1b}) and (\ref{eq:2.5a})-(\ref{eq:2.5g}), 
the asymptotic form of the energy density (\ref{eq:1.3}) is 

\begin{equation}
\varepsilon=\frac{3}{4\kappa}\frac{1}{t^2}+\frac{3}{2\kappa}\frac{k}{S_1}
\frac{1}{t}+\frac{1}{\kappa}\Bigl(\mu^2-\frac{\Lambda}{3}+\frac{9}{4}
\frac{k^2}{S^2_1}\Bigr)+{\cal O}(t). \label{eq:2.7}
\end{equation}
Hence (\ref{eq:2.6}) is satisfied identically in the ${\cal O}(t^{-2})$, 
${\cal O}(t^{-1})$ and ${\cal O}(1)$ orders. We expect that (\ref{eq:2.4}) is 
satisfied identically in any order, and hence, in particular, (\ref{eq:2.6}) 
provides the energy density even with ${\cal O}(t)$ accuracy. Indeed, using 
{\tt Mathematica}, we found that (\ref{eq:2.6}) is satisfied identically even 
in the ${\cal O}(t^8)$ order with the ${\cal O}(t^8)$ accurate solutions of 
the other two field equations. For $k=0$ the expansion coefficients $\phi_m$ 
do not depend on $S_1$, while the non-zero expansion coefficients $S_n$ are 
all proportional to $S_1$ even in the ${\cal O}(t^8)$ accurate solutions. 
Hence, in the solutions for $k=0$ the parameter $S_1$ plays the role only of 
a physically irrelevant overall scale factor. 

Therefore, (\ref{eq:2.1a})-(\ref{eq:2.1b}) with 
(\ref{eq:2.5a})-(\ref{eq:2.5g}) provide the asymptotic solution of the 
field equations that is singular at $\Phi^2=6/\kappa$. The only freely 
specifiable initial datum is $S_1$ (and the discrete parameter $k$). Since 
$h_{ab}={}_1h_{ab}(S_1t-kt^2+\frac{1}{3}(\mu^2+\frac{2}{3}\Lambda)S_1t^3+{\cal 
O}(t^4))$, this specifies the intrinsic 3-geometry of the hypersurfaces 
$\Sigma_t$. Since we want \emph{real} scale function $S(t)$ for any $t>0$, 
the coefficient $S_1$ must be positive. Hence, this solution cannot be 
extended to the domain $t<0$, where $S(t)$ would be imaginary. By 

\begin{equation}
\chi=3\frac{\dot S}{S}=\frac{3}{2}\frac{1}{t}\Bigl(1-\frac{k}{S_1}t+\bigl(
\frac{2}{3}(\mu^2+\frac{2}{3}\Lambda)-\frac{k^2}{S^2_1}\bigr)t^2+{\cal O}
(t^3)\Bigr) \label{eq:2.8}
\end{equation}
and (\ref{eq:2.7}), near the singularity, both the mean curvature and the 
energy density have a \emph{universal} character: They do not depend on the 
initial datum $S_1$ in the leading order, and not even on the parameters 
$\mu^2$ and $\lambda$ of the Higgs sector in the first two orders. Since 
by (\ref{eq:2.7}) and $3P=\varepsilon-\mu^2\Phi^2$ we have that 

\begin{equation}
\bigl(R_{ab}-\frac{1}{2}Rg_{ab}+\Lambda g_{ab}\bigr)\bigl(R^{ab}-\frac{1}{2}R
g^{ab}+\Lambda g^{ab}\bigr)=\kappa^2T_{ab}T^{ab}=\kappa^2(\varepsilon^2+3P^2)
\sim t^{-4}, \label{eq:2.E}
\end{equation}
the singularity at $t=0$ is a \emph{physical, scalar polynomial curvature 
singularity of the spacetime} (see \cite{HE}). Since the Higgs field and 
the curvature scalar remains finite, $\Phi^2\rightarrow 6/\kappa$ and $R
\rightarrow 4\Lambda+6\mu^2<0$ as $t\rightarrow0$, this singularity is 
`weaker' then the Big Bang of subsection \ref{sub-3.2} (in which $R$, 
$R_{ab}R^{ab}$ and $\Phi^2$ are all diverging). Thus, we call this the Small 
Bang singularity. 

Since by (\ref{eq:2.8}) $\chi$ is strictly monotonically decreasing, this 
can in fact be used as a natural time coordinate (`York time') in a 
neighbourhood of the singularity. By (\ref{eq:2.8}) the \emph{proper time} 
corresponding to the hypersurface $\Sigma_t$ with the critical value $\chi_c$ 
of the mean curvature, i.e. to the instant of the `genesis' of the rest 
masses and electric charge via the BEH mechanism, is $t_c\simeq3/(2\chi_c)$. 
For earlier times the rest mass of the Higgs field is not defined. 

Since $\chi^2_c:=9(\mu^2+\Lambda/3+3\lambda/\kappa)>0$, by (\ref{eq:2.1a}), 
(\ref{eq:2.5d}) and $\phi_2<0$ (see (\ref{eq:2.5e})) the Higgs field locally 
takes its \emph{maximal} value at $t=0$; i.e. this solution \emph{cannot be 
continued} to the $\Phi^2>6/\kappa$ side of the $\Phi=\sqrt{6/\kappa}$ line 
in the configuration space. Moreover, even though $S^2(t)$ with $S_1<0$ for 
$t<0$ appears to be a solution reaching the $\Phi=\sqrt{6/\kappa}$ line from 
the $\Phi^2>6/\kappa$ side of the configuration space, it \emph{cannot} be a 
solution. In fact, if $(\Phi(t),S(t))$ were such a solution which approached 
the $\Phi=\sqrt{6/\kappa}$ line form the $\Phi^2>6/\kappa$ side of the 
configuration space, then the Higgs field would take its \emph{minimum} on 
the $\Phi=\sqrt{6/\kappa}$ line, which would contradict $\phi_1=0$ and $\phi
_2<0$. Therefore, this line, \emph{as a singularity}, cannot be reached by 
solutions with asymptotics $S^2(t)={\cal O}(t)$ from the $\Phi^2>6/\kappa$ 
part of the configuration space. 

To summarize, the field equations have asymptotic power series solutions in 
which the scale function is $S^2={\cal O}(t)$, the norm of the Higgs field 
$\vert\Phi\vert$ tends to its \emph{maximal} value, $\sqrt{6/\kappa}$, as 
${\cal O}(t^2)$, and the energy density of the Higgs field is 
\emph{diverging}. For $k=\pm1$ the solutions depend on a positive, freely 
specifiable parameter, viz. $S_1$, but for $k=0$ this parameter plays the 
role only as an overall scale factor. The singularity is a physical, 
\emph{scalar polynomial curvature singularity of the spacetime}, though the 
curvature scalar remains bounded. Thus we call this the Small Bang 
singularity. This singularity can be reached by power series type asymptotic 
solutions only from the $\Phi^2<6/\kappa$ side of the configuration space. 
In the vicinity of the initial singularity the EccH system does not have any 
instantaneous vacuum state, and hence then the rest mass of the Higgs field 
is not defined.

\subsection{An exceptional solution with a Milne type singularity}
\label{sub-2.3}

Next, let us suppose that $S_0=S_1=0$ and $S_2\not=0$ (i.e. $\phi_0=1$ and 
$n=2$ in (\ref{eq:2.2})-(\ref{eq:2.4})). Then, repeating the analysis of 
the previous subsection, we find that $k\not=0$ and 

\begin{eqnarray}
&{}&S_2=-k, \hskip 10pt 
  S_3=S_5=0, \hskip 10pt 
  S_4=-\frac{k}{6}\bigl(\mu^2+\frac{2}{3}\Lambda\bigr); \hskip 10pt 
  \phi_1=\phi_3=0, \label{eq:2.9a} \\
&{}&\phi_2=-\frac{1}{4}\bigl(\mu^2+\frac{\Lambda}{3}+3\frac{\lambda}{\kappa}
  \bigr), \hskip 10pt
  \phi_4=\frac{1}{96}\bigl(5\mu^2+\frac{4}{3}\Lambda+18\frac{\lambda}{\kappa}
  \bigr)(\mu^2+\frac{\Lambda}{3}+3\frac{\lambda}{\kappa}). \label{eq:2.9b}
\end{eqnarray}
The scale function $S(t)$ can be real for any $t>0$ only if $k=-1$. Hence, 
this asymptotic solution, being an \emph{even} function of time, is time 
symmetric (with respect to the $t=0$ hypersurface), and, by $\phi_2<0$, the 
Higgs field takes its \emph{maximal} value at $t=0$. Thus, in particular, the 
dynamical trajectory corresponding to this solution in the configuration 
space only touches, but does not cross the $\Phi=\sqrt{6/\kappa}$ line. Also, 
this line cannot be reached by a solution with asymptotics $S^2(t)={\cal O}
(t^2)$ from the $\Phi^2>6/\kappa$ side of the configuration space. This 
solution appears to be exceptional in the sense that it is \emph{uniquely 
determined} by the parameters $\Lambda$, $\kappa$, $\mu^2$ and $\lambda$ of 
the EccH model and does not depend on any freely specifiable initial 
condition. However, as we will see in subsection \ref{sub-2.5}, this solution 
belongs to a whole 1-parameter family of solutions. 

By $S^2(t)=t^2(1+\frac{1}{6}(\mu^2+\frac{2}{3}\Lambda)t^2+{\cal O}(t^4))$ the 
curvature scalar of the intrinsic metric and the mean curvature of the 
hypersurfaces $\Sigma_t$, respectively, are 

\begin{equation*}
{\cal R}=-\frac{6}{t^2}\Bigl(1-\frac{1}{6}\bigl(\mu^2+\frac{2}{3}\Lambda
\bigr)t^2+{\cal O}(t^4)\Bigr), \hskip 20pt
\chi=\frac{3}{t}\Bigl(1+\frac{1}{6}\bigl(\mu^2+\frac{2}{3}\Lambda
\bigr)t^2+{\cal O}(t^4)\Bigr);
\end{equation*}
and hence the singularity has a universal character, independently of the 
parameters $\mu^2$ and $\lambda$ of the Higgs sector. Since the mean curvature 
is diverging as $t\rightarrow 0$, the rest mass of the Higgs field cannot be 
defined on the time interval $(0,t_c)$ for some $t_c>0$. 

In this solution the energy density is \emph{bounded}: 

\begin{equation}
\varepsilon=\frac{3}{2\kappa}\mu^2+{\cal O}(t^2). \label{eq:2.10}
\end{equation}
At the singularity $t=0$ it is not only the spacetime curvature scalar, but 
also the whole Ricci tensor remains finite. In fact, by Einstein's equations, 
$3P=\varepsilon-\mu^2\Phi^2$ and equation (\ref{eq:2.10}) it is $R_{ab}=t_a
t_b(\Lambda+\frac{1}{2}\kappa\mu^2\Phi^2-\kappa\varepsilon)+h_{ab}(\Lambda+
\frac{1}{2}\kappa\mu^2\Phi^2+\kappa P)=(\Lambda+\frac{3}{2}\mu^2+{\cal O}
(t^2))g_{ab}$. We show that the singularity in this solution is analogous to 
that in the Milne universe \cite{ES}, i.e. it corresponds to a \emph{regular 
boundary point} of the spacetime through which the solution can be extended 
into a larger spacetime manifold. 

To see this, let us recall that the Milne universe is a special FRW spacetime 
(e.g. with the line element $ds^2=dt^2-S^2(t)(d\rho^2+\sinh^2\rho(d\theta^2+
\sin^2\theta d\phi^2))$ with the coordinate ranges $t>0$, $\rho\geq0$ and 
$(\theta,\phi)\in S^2$) in which the scale function is $S(t)=t$. In the new 
coordinates $\tau:=t\cosh\rho$, $r:=t\sinh\rho$ the Milne universe turns out 
to be the $\tau>r\geq0$ part, i.e. just the chronological future of the 
origin, of the Minkowski spacetime \cite{ES}. Since in our solution the scale 
function deviates from that of the Milne universe only in higher order terms, 
viz. $S^2(t)=t^2(1+\frac{1}{6}(\mu^2+\frac{2}{3}\Lambda)t^2+{\cal O}(t^4))$, 
it seems natural to introduce the new coordinates analogously: $\tau:=S(t)
\cosh\rho$ and $r:=S(t)\sinh\rho$. The range of these coordinates is $\tau
>0$ and $r\geq0$; and the singularity $t=0$ of the solution corresponds to 
$\tau=0$. In these coordinates the line element is 

\begin{eqnarray*}
ds^2\!\!\!\!&=\!\!\!\!&\Bigl(\frac{1}{\dot S^2}\cosh^2\rho-\sinh^2\rho\Bigr)
  d\tau^2-2\sinh\rho\cosh\rho\Bigl(\frac{1}{\dot S^2}-1\Bigr)\,d\tau\, dr- \\
&{}&-\Bigl(\cosh^2\rho-\frac{1}{\dot S^2}\sinh^2\rho\Bigr)dr^2-
  r^2\bigl(d\theta^2+\sin^2\theta \,d\phi^2)= \\
\!\!\!\!&=\!\!\!\!&d\tau^2-dr^2-r^2\bigl(d\theta^2+\sin^2\theta\,d\phi^2\bigr)
  -\\
 &{}&-\bigl(\frac{1}{2}(\mu^2+\frac{2}{3}\Lambda)t^2+{\cal O}(t^4)\bigr)\Bigl(
  \cosh^2\rho \,d\tau^2-2\cosh\rho\sinh\rho \,d\tau \,dr+\sinh^2\rho \,dr^2
  \Bigr),
\end{eqnarray*}
where now $t$ and $\rho$ are considered to be functions of $\tau$ and $r$. 
Clearly, this line element is perfectly regular even at $\tau=0$ (when $t=
0$), and the range of the new time coordinate certainly can be extended 
to zero and even to negative values. Therefore, the singularity of the 
solution at $t=0$ is a singularity of the \emph{foliation} $\Sigma_t$ only, 
but \emph{not} of the spacetime itself. Since this solution is an \emph{even} 
function of $t$, it is well defined for $t<0$. Hence, in the leading order, 
it describes an evolution of the EccH system in which, near $\tau=0$ for 
$\tau<0$, the Higgs field is increasing; the system reaches the regular state 
at $\tau=0$ in which $\Phi^2=6/\kappa$ and when it `bounces back'; and then 
it continues its evolution (for $\tau>0$) in the $\Phi^2<6/\kappa$ regime in 
which the Higgs field is decreasing. At the instant of the `bounce' the 
foliation $\Sigma_t$ becomes singular. However, taking into account the next 
order correction, by $S_4=(\mu^2+2\Lambda/3)/6<0$, the scale function $S^2
(t)$ has \emph{local maximum} at $t^2_m\simeq-3(\mu^2+2\Lambda/3)$. 
Nevertheless, the large scale behaviour of the solution can be revealed only 
by numerics. 

As we already noted, in subsection \ref{sub-2.5} we will see that this 
exceptional solution can be considered as a member of a 1-parameter family of 
asymptotic solutions. 

Finally, in the rest of this subsection, we show that there is no more 
asymptotic power series solution in the $\Phi^2\rightarrow 6/\kappa$, $S^2
\rightarrow 0$ case. Thus, first, let us suppose that $S_0=S_1=...=S_{n-1}=0$ 
and $S_n\not=0$ for some $n\geq3$. Then a straightforward calculation shows 
that this assumption on the structure of the scale function contradicts 
equation (\ref{eq:2.3}); i.e. there is no asymptotic solution of the field 
equations with this structure. Similarly, if $0<n<1$ or $1<n<2$, then the 
leading order term in equation (\ref{eq:2.3}) would be only $n(n-1)t^{-2}$, 
which cannot be vanishing. If $n\in(2,\infty)-\mathbb{N}$, then the leading 
order term in (\ref{eq:2.3}) would be $(2k/S_n)t^{-n}$, whose vanishing would 
imply $k=0$. Substituting this back into (\ref{eq:2.3}) the leading order 
term in the resulting equation would be $n(n-1)t^{-2}$ again, yielding a 
contradiction.

\subsection{A family of regular asymptotic solutions}
\label{sub-2.4}

Now let us still suppose that $\phi_0=1$, but assume that $S_0>0$; i.e. 
although the energy density is \emph{formally} singular, but the spacetime 
geometry is not. Now we show that this singularity of $\varepsilon$ is only 
\emph{fictitious}, and the solution is completely regular. Then equations 
(\ref{eq:2.2})-(\ref{eq:2.3}) (with $\phi_0=1$ and $n=0$) yield 

\begin{eqnarray}
&{}&\frac{S_2}{S_0}=-\frac{k}{S_0}+(\mu^2+\frac{2}{3}\Lambda), 
  \label{eq:2.11a} \\
&{}&\frac{S_3}{S_0}=\frac{1}{3}(\mu^2+\frac{2}{3}\Lambda)\frac{S_1}{S_0}+
  \frac{2}{3}\mu^2\phi_1, \label{eq:2.11b} \\
&{}&\phi_2=-\frac{3}{4}\phi_1\frac{S_1}{S_0}-(\mu^2+\frac{\Lambda}{3}+3
  \frac{\lambda}{\kappa}), \label{eq:2.11c} \\
&{}&\phi_3=\frac{1}{2}(\mu^2+\frac{\Lambda}{3}+3\frac{\lambda}{\kappa})
  \frac{S_1}{S_0}+\frac{1}{6}\phi_1\Bigl(3\frac{k}{S_0}+\frac{15}{4}
  \frac{S^2_1}{S^2_0}-7\mu^2-\frac{8}{3}\Lambda-18\frac{\lambda}{\kappa}
  \Bigr), \label{eq:2.11d} \\
&{}&\phi_4=-\frac{1}{2}\phi^2_1(\mu^2+3\frac{\lambda}{\kappa})+\phi_1
  \frac{S_1}{S_0}\Bigl(-\frac{35}{64}\frac{S^2_1}{S^2_0}-\frac{15}{16}
  \frac{k}{S_0}+\frac{21}{16}\mu^2+\frac{5}{8}\Lambda+\frac{9}{4}
  \frac{\lambda}{\kappa}\Bigr)+ \nonumber \\
&{}&\hskip 26pt +(\mu^2+\frac{\Lambda}{3}+3\frac{\lambda}{\kappa})\Bigl(
  -\frac{7}{16}\frac{S^2_1}{S^2_0}-\frac{1}{2}\frac{k}{S_0}+\frac{5}{6}\mu^2+
  \frac{7}{18}\Lambda+\frac{3}{2}\frac{\lambda}{\kappa}\Bigr). \label{eq:2.11e}
\end{eqnarray}
Thus, $S_0$, $S_1$, $\phi_1$ and the discrete parameter $k$ determine $\Phi$ 
and $S^2$ up to order ${\cal O}(t^4)$ and ${\cal O}(t^3)$, respectively, 
\emph{provided} (\ref{eq:2.4}), the difference of the two Einstein equations, 
is satisfied. Using (\ref{eq:2.11a}) and (\ref{eq:2.11b}), equation 
(\ref{eq:2.4}) takes the form 

\begin{equation}
\frac{4\kappa}{3}\varepsilon=\frac{S^2_1}{S^2_0}-\frac{4}{3}\Lambda+\frac{4k}
{S_0}-2\frac{S_1}{S_0}\Bigl(\frac{S^2_1}{S^2_0}-\frac{4}{3}\Lambda+\frac{4k}
{S_0}-2\mu^2\Bigr)t+{\cal O}(t^2). \label{eq:2.12}
\end{equation}
To evaluate this, we need the asymptotic expansion of the energy density 
(\ref{eq:1.3}). 

Thus, first suppose that $\phi_1\not=0$. Then the leading term in the 
expansion of $\varepsilon$ will be 

\begin{equation*}
\frac{2}{3}\kappa\varepsilon=-\frac{1}{t}\frac{1}{\phi_1}\Bigl(\phi^2_1+
\frac{S_1}{S_0}\phi_1+\mu^2+\frac{\Lambda}{3}+3\frac{\lambda}{\kappa}\Bigr)
+{\cal O}(1).
\end{equation*}
Substituting this into (\ref{eq:2.12}), we find that this leading order term 
in $\varepsilon$ must be zero: 

\begin{equation}
0=\phi^2_1+\frac{S_1}{S_0}\phi_1+\mu^2+\frac{\Lambda}{3}+3\frac{\lambda}
{\kappa}. \label{eq:2.13}
\end{equation}
Then, it is a lengthy but straightforward calculation to check that 
(\ref{eq:2.12}), with the ${\cal O}(t)$ accurate expansion of the energy 
density, is already satisfied identically even in the next two (i.e. 
${\cal O}(1)$ and ${\cal O}(t)$) orders, and then (\ref{eq:2.12}) becomes 
the ${\cal O}(t)$ accurate expression of the energy density. In particular, 
(\ref{eq:2.12}) shows that the energy density is \emph{bounded}. By 
equation (\ref{eq:2.13}), the expansion coefficients in $\Phi$ with ${\cal 
O}(t^4)$ accuracy are 

\begin{eqnarray}
\phi_1\!\!\!\!&=\!\!\!\!&-\frac{1}{2}\frac{S_1}{S_0}\pm\frac{1}{2}\sqrt{(
  \frac{S_1}{S_0})^2-4(\mu^2+\frac{\Lambda}{3}+3\frac{\lambda}{\kappa})}, 
  \label{eq:2.14a} \\
\phi_2\!\!\!\!&=\!\!\!\!&\phi_1\bigl(\phi_1+\frac{1}{4}\frac{S_1}{S_0}\bigr), 
  \label{eq:2.14b} \\
\phi_3\!\!\!\!&=\!\!\!\!&-\frac{1}{2}\phi_1\Bigl(\frac{S_1}{S_0}\phi_1-
  \frac{1}{4}(\frac{S_1}{S_0})^2+\frac{7}{3}\mu^2+\frac{8}{9}\Lambda+6
  \frac{\lambda}{\kappa}-\frac{k}{S_0}\Bigr), \label{eq:2.14c} \\
\phi_4\!\!\!\!&=\!\!\!\!&\frac{1}{4}\phi_1\frac{S_1}{S_0}\Bigl(-\frac{7}{4}
  \frac{k}{S_0}-\frac{7}{16}\frac{S^2_1}{S^2_0}+\frac{23}{12}\mu^2+\frac{17}
  {18}\Lambda+3\frac{\lambda}{\kappa}\Bigr)- \nonumber \\
\!\!\!\!&{}\!\!\!\!&-\frac{1}{2}\phi^2_1\Bigl(-\frac{k}{S_0}-\frac{7}{8}
  \frac{S^2_1}{S^2_0}+\frac{8}{3}\mu^2+\frac{7}{9}\Lambda+6\frac{\lambda}
  {\kappa}\Bigr). \label{eq:2.14d}
\end{eqnarray}
To ensure the reality of $\phi_1$, the mean curvature at $t=0$ cannot be 
smaller than its critical value: $\chi^2(0)=\frac{9}{4}(S_1/S_0)^2\geq9
(\mu^2+\Lambda/3+3\lambda/\kappa)=:\chi^2_c$. Since the canonical momentum 
$\Pi$ at $t=0$ is 

\begin{equation}
\Pi(0)=\sqrt{\frac{6}{\kappa}}(\phi_1+\frac{1}{2}\frac{S_1}{S_0})=\pm
\frac{1}{2}\sqrt{\frac{6}{\kappa}}\sqrt{(\frac{S_1}{S_0})^2-4(\mu^2+
\frac{\Lambda}{3}+3\frac{\lambda}{\kappa})}=\pm\sqrt{\frac{2}{3\kappa}}
\sqrt{\chi^2(0)-\chi^2_c}, \label{eq:2.pi}
\end{equation}
this condition is equivalent to the reality of $\Pi(0)$, too. Therefore, the 
freely specifiable parameters of the solution are 

\begin{equation}
 S_0>0, \qquad  \vert\frac{S_1}{S_0}\vert\geq\frac{2}{3}\chi_c, \label{eq:reg}
\end{equation}
the discrete parameter $k$ and the sign of $\Pi(0)$. By means of these the 
initial value of the basic canonical field variables at $t=0$ can be given: 
$\Phi(0)=\sqrt{6/\kappa}$, the $\Pi(0)$ given above, $\chi(0)=\frac{3}{2}
S_1/S_0$ and $S^2(0)=S_0$. The smallest allowed value for $\chi^2(0)$, viz. 
$\chi^2_c$, corresponds to $\Pi(0)=0$; and $S_1>0$ corresponds to 
\emph{expansion} of the universe at $t=0$. With these conditions on $S_1$ 
and the mean curvature $\chi(0)$ the expansion coefficient $\phi_1$ is 
always \emph{negative} for any sign in (\ref{eq:2.14a}). Thus, the present 
regular solutions \emph{can} be continued for $t<0$ in the $\Phi^2>6/\kappa$ 
domain of the configuration space. 

If $\chi^2(0)>\chi^2_c$, then, on the interval $[0,t_c)$ for some $t_c>0$, 
the mean curvature $\chi$ is certainly greater than $\chi_c$. Thus, on this 
interval, the EccH system does \emph{not} have any instantaneous vacuum 
state, and hence the rest mass of the Higgs field cannot be defined. 

Next suppose that $\chi^2(0)=\chi^2_c$. In this case the mean curvature is 

\begin{equation}
\chi=\chi_c-3\Bigl(\mu^2+6\frac{\lambda}{\kappa}+\frac{k}{S_0}\Bigr)t+\chi_c
\Bigl(\mu^2+12\frac{\lambda}{\kappa}+3\frac{k}{S_0}\Bigr)t^2+{\cal O}(t^3).
\label{eq:2.15}
\end{equation}
Since $\mu^2+6\lambda/\kappa>0$ holds, $\dot\chi(0)<0$ follows for $k=0,1$ 
and for any value of $S_0$. Hence, for $k=0,1$, the rest mass of the Higgs 
field cannot be defined on the time interval $(-t_c,0]$ for some $t_c>0$, 
and the BEH mechanism could start to work only \emph{after} $t=0$. If $k=-1$, 
then $\dot\chi(0)<0/=0/>0$ holds precisely when $S_0>S^2_c/=S^2_c/<S^2_c$, 
respectively, where 

\begin{equation}
S^2_c:=\Bigl(\mu^2+6\frac{\lambda}{\kappa}\Bigr)^{-1}\simeq(10^{-32}cm)^2. 
\label{eq:2.16}
\end{equation}
(Interestingly enough, $S^2_c$ is just the $\chi\rightarrow\pm\chi_c$ limit of 
the scale function $S^2_v(\chi)$ at the global instantaneous gauge symmetry 
breaking vacuum states given by (\ref{eq:1.5}).) Thus, for $S_0\in(0,S^2_c)$ 
the mean curvature is definitely greater than $\chi_c$ on $[0,t_c)$ for some 
$t_c>0$, and hence the BEH mechanism could start to work only \emph{after} 
$t=t_c$. If $S_0\in(S^2_c,\infty)$, then the mean curvature exceeds $\chi_c$ 
on some interval $(-t_c,0]$, $t_c>0$, and the BEH mechanism starts to work 
only \emph{after} $t=0$. 

On the other hand, since $\mu^2+3\lambda/\kappa>0$, in the special solution 
with $S_0=S^2_c$ one has $\ddot\chi(0)=-4\chi_c(\mu^2+3\lambda/\kappa)<0$, 
i.e. the mean curvature takes its \emph{maximal} value, $\chi_c$, at $t=0$. 
Therefore, in the single, exceptional asymptotic power series solution with 
the initial data $S_0=S^2_c$, $S_1/S_0=2\chi_c/3$ and $k=-1$ the criterion 
$\chi^2<\chi^2_c$ of the existence of instantaneous vacuum states is 
satisfied on a time interval around $t=0$ \emph{except} only the instant 
$t=0$. 

Next, suppose that in the expansion (\ref{eq:2.1a}) $\phi_1=0$ but $\phi_2
\not=0$. Then (\ref{eq:2.2}) yields $\phi_2=-(\mu^2+\Lambda/3+3\lambda/
\kappa)$, the energy density (\ref{eq:1.3}) has the structure 

\begin{equation*}
\frac{2}{3}\kappa\varepsilon=-\frac{1}{t^2}\frac{1}{\phi_2}\bigl(\mu^2+
\frac{\Lambda}{3}+3\frac{\lambda}{\kappa}\bigr)+{\cal O}(t^{-1}),
\end{equation*}
while (\ref{eq:2.3}) still gives $S_2=-k+(\mu^2+2\Lambda/3)S_0$. Using these, 
(\ref{eq:2.4}) yields $(\mu^2+\Lambda/3+3\lambda/\kappa)/\phi_2=0$, which is 
a contradiction. Finally, suppose that $\phi_1=\phi_2=...=\phi_{k-1}=0$, but 
$\phi_k\not=0$ for some $k=3,4,...$. Then (\ref{eq:2.2}) gives $\mu^2+
\Lambda/3+3\lambda/\kappa=0$ and $\phi_k=0$, which is a contradiction. 
Therefore, in addition to the solution given by 
(\ref{eq:2.11a})-(\ref{eq:2.11b}) and (\ref{eq:2.13})-(\ref{eq:2.14d}), 
there is no more asymptotic power series solution with $S_0>0$, $\phi_0=1$. 

To summarize, according to the two signs in (\ref{eq:2.14a}), there are two 
\emph{regular} 2-parameter families of asymptotic power series solutions of 
the field equations through the $\Phi=\sqrt{6/\kappa}$ line of the 
configuration space for $k=0,\pm1$. The two parameters, $S_0$ and $S_1$, 
are subject to the inequalities $S_0>0$ and $(S_1/S_0)^2\geq4\chi^2_c/9$. A 
solution represents an expanding universe at $t=0$ precisely when $S_1>0$. 
The significance of these regular solutions is that, although formally the 
configuration $\Phi^2=6/\kappa$ is a singularity of the energy density, 
there could be solutions describing a large expanding universe at the late, 
weakly gravitating era but with Big Bang type initial singularity: The 
$\Phi^2=6/\kappa$ hypersurface in spacetime could be `traversable'. Apart 
from an exceptional, regular solution (with specific parameters), in all 
these regular solutions there is an open time interval (before or after the 
instant $t=0$) on which the EccH system does not have any instantaneous 
vacuum state; while in the exceptional regular solution it is only the 
instant $t=0$.

\subsection{A family of solutions with a Milne type singularity}
\label{sub-2.5}

Finally, let us suppose that the Higgs field is still bounded and $S^2(t)
\rightarrow 0$, but $\Phi^2(t)\not\rightarrow 6/\kappa$ as $t\rightarrow0$. 
Thus in the expansion of the scale function $S_0=0$ but $S_n\not=0$ for 
some $n>0$, and in the expansion of the Higgs field $\phi^2_0\not=1$. The 
field equation (\ref{eq:2.2}) for the Higgs field (with $\phi^2_0\not=1$) 
yields 

\begin{eqnarray}
\phi_1\!\!\!\!&=\!\!\!\!&0, \label{eq:2.17a} \\
(2+3n)\phi_2\!\!\!\!&=\!\!\!\!&-\phi_0\Bigl(\mu^2+\frac{2}{3}\Lambda+\bigl(
  \mu^2+6\frac{\lambda}{\kappa}\bigr)\phi^2_0\Bigr), \label{eq:2.17b} \\
(2+\frac{3}{2}n)\phi_3\!\!\!\!&=\!\!\!\!&-\frac{S_{n+1}}{S_n}\phi_2, 
  \nonumber \\
6(2+n)\phi_4\!\!\!\!&=\!\!\!\!&-\Bigl(6\frac{S_{n+2}}{S_n}-3\frac{7+3n}{4+3n}
  (\frac{S_{n+1}}{S_n})^2+\mu^2+\frac{2}{3}\Lambda+3\bigl(\mu^2+6
  \frac{\lambda}{\kappa}\bigr)\phi^2_0\Bigr)\phi_2. \nonumber
\end{eqnarray}
Then by (\ref{eq:2.17a}) and (\ref{eq:2.17b}) the energy density is 

\begin{equation*}
\kappa\varepsilon=\frac{6n}{2+3n}\mu^2\phi^2_0+3\frac{2-n}{2+3n}
\frac{\phi^2_0}{1-\phi^2_0}\bigl(\mu^2+\frac{\Lambda}{3}+3\frac{\lambda}
{\kappa}\phi^2_0\bigr)+{\cal O}(t).
\end{equation*}
Hence, by (\ref{eq:2.4}), $n=2$ must hold. In fact, if $n$ were greater 
than $2$, then the leading order term in (\ref{eq:2.4}) would be $(2k/S_n)
t^{-n}$, whose vanishing would imply $k=0$. However, substituting $k=0$ back 
into (\ref{eq:2.4}) the leading term would be $n/t^2$, which cannot be zero. 
On the other hand, if $2>n>0$ held, then the leading order term would also 
be $n/t^2$, which cannot be vanishing. 

With the substitution $n=2$ and $\phi_1=0$ equation (\ref{eq:2.3}) yields 
that $k\not=0$ and 

\begin{equation}
S_2=-k, \hskip 20pt
S_3=0, \hskip 20pt
S_4=-\frac{k}{6}\bigl(\frac{2}{3}\Lambda+\mu^2\phi^2_0\bigr), \hskip 20pt
S_5=0; \label{eq:2.18a}
\end{equation}
by means of which 

\begin{equation}
\phi_3=0, \hskip 20pt
\phi_4=\frac{1}{192}\Bigl(\mu^2+\frac{4}{3}\Lambda+4\mu^2\phi^2_0+18
\frac{\lambda}{\kappa}\phi^2_0\Bigr)\bigl(\mu^2+\frac{2}{3}\Lambda+(\mu^2+
6\frac{\lambda}{\kappa})\phi^2_0\bigr)\phi_0. \label{eq:2.18b}
\end{equation}
Since $S^2(t)$ must be positive for any $t>0$, $k=-1$ must hold. With these 
substitutions the energy density takes the form 

\begin{equation}
\varepsilon=\frac{3}{2\kappa}\mu^2\phi^2_0+{\cal O}(t^2). \label{eq:2.19}
\end{equation}
Calculating the energy density with ${\cal O}(t^4)$ accuracy and substituting 
the resulting expression to the field equation (\ref{eq:2.4}) we find that it 
is satisfied identically. 

Thus, (\ref{eq:2.17b})-(\ref{eq:2.18b}) provide a 1-parameter family of 
solutions of the field equations. The solution is an \emph{even} function of 
the cosmological proper time with scale function $S^2(t)=t^2(1+{\cal O}
(t^2))$, and the domain of the parameter $\phi_0$ is the union of the 
disjoint intervals $(-\infty,-1)$, $(-1,1)$ and $(1,\infty)$. Because of the 
$\mathbb{Z}_2:\Phi\mapsto-\Phi$ gauge symmetry, it is enough to consider 
only the domain $[0,1)\cup(1,\infty)$ for $\phi_0$. Comparing the expansion 
coefficients (\ref{eq:2.17a})-(\ref{eq:2.18b}) with those in 
(\ref{eq:2.9a})-(\ref{eq:2.9b}), and also (\ref{eq:2.19}) with 
(\ref{eq:2.10}), we find that the exceptional solution found in subsection 
\ref{sub-2.3} fits naturally into the present 1-parameter family of solutions 
with parameter value $\phi^2_0=1$. 

However, because of the presence of the free parameter $\phi_0$, the 
qualitative properties of the scale function $S^2(t)$ of the present 
1-parameter family of solutions can be even more diverse than that of the 
exceptional solution of subsection \ref{sub-2.3}. In particular, the 
expansion coefficient $S_4$ is not necessarily negative: That could be 
zero or can have any sign. 

Also, the behaviour of the Higgs field could be even qualitatively different 
from that of the exceptional solution of subsection \ref{sub-2.3}. First, for 
$\phi_0=0$ the Higgs field is vanishing; in which case the expansion 
coefficients of the scale function are just those of the de Sitter spacetime. 
Thus, for $\phi_0=0$, the solution could be expected to be locally isometric 
to the de Sitter spacetime. If $\phi_0\not=0$, then by $\phi_1=0$ the 
singularity at $t=0$ is still a critical point of the Higgs field. By 
(\ref{eq:2.17b}) this is a local maximum precisely when 

\begin{equation}
\phi^2_0>(\phi^{crit}_0)^2:=-\frac{\kappa}{6}\frac{\mu^2+\frac{2}{3}\Lambda}
{\lambda+\frac{\kappa}{6}\mu^2}\simeq 1.8\times10^{-33}. \label{eq:2.20}
\end{equation}
The corresponding maximal value is not bounded from above, that can even be 
greater than $\sqrt{6/\kappa}$; while the infimum of these maximal values 
is $\phi^{crit}_0\sqrt{6/\kappa}\simeq\sqrt{-\mu^2/\lambda}$, just the vacuum 
value of the Higgs field of the Weinberg--Salam model in Minkowski spacetime. 
If $\phi_0=\phi^{crit}_0$, then $\Phi$ is constant in the ${\cal O}(t^4)$ 
order, while for $\phi_0<\phi^{crit}_0$ it is has a local \emph{minimum} at 
$t=0$. 

To summarize, the field equations have a 1-parameter family of asymptotic 
power series singular solutions in which the scale function is $S^2(t)=
{\cal O}(t^2)$, the Higgs field is bounded, the parameter of the solutions 
is just the value of the Higgs field at $t=0$, and this cosmological model 
is necessarily the hyperboloidal one ($k=-1$). However, the singularity is 
a singularity only of the \emph{foliation} of the spacetime, but the 
spacetime itself is not singular. These solutions are \emph{even} functions 
of $t$, and hence they can be extended to $t\leq0$. In this form, near $t=0$, 
the solutions describe contracting, and then expanding universes. In an early 
period just after the `bouncing' at $t=0$, the Higgs sector does not have any 
instantaneous vacuum state.


\section{Asymptotic solutions with diverging Higgs field}
\label{sec:3}

In this section we determine \emph{all} the asymptotic power series 
solutions of the field equations in which the Higgs field is diverging, 
$\Phi^2(t)\rightarrow\infty$ if, say, $t\rightarrow0$. Since then the energy 
density is also diverging, the geometry is necessarily singular: $S^2(t)
\rightarrow0$. Thus, let us suppose that this singularity is reached at 
$t=0$, and we can expand these fields as 

\begin{eqnarray}
\Phi&\!\!\!\!=&\!\!\!\!\sqrt{\frac{6}{\kappa}}\Bigl(\phi_{-m}t^{-m}+\phi_{-m+1}
  t^{-m+1}+\phi_{-m+2}t^{-m+2}+\phi_{-m+3}t^{-m+3}+{\cal O}(t^{-m+4})\Bigr), 
  \label{eq:3.1a} \\
S^2&\!\!\!\!=&\!\!\!\!S_nt^n+S_{n+1}t^{n+1}+S_{n+2}t^{n+2}+S_{n+3}t^{n+3}+{\cal O}
  (t^{n+4}) 
  \label{eq:3.1b}
\end{eqnarray}
for some $n,m>0$ and real constant coefficients $\phi_{-m},\phi_{-m+1},...$ 
and $S_{n},S_{n+1},...$ with $\phi_{-m}\not=0$ and $S_n>0$.

\subsection{The field equations}
\label{sub-3.1}

Substituting these expansions into the evolution equation (\ref{eq:1.2}) for 
the Higgs field, we obtain 

\begin{eqnarray}
0\!\!\!\!&=\!\!\!\!&m(m+1-\frac{3}{2}n)\phi_{-m}t^{-m-2}+\Bigl((m-1)(m-
  \frac{3}{2}n)\phi_{-m+1}-\frac{3}{2}m\frac{S_{n+1}}{S_n}\phi_{-m}\Bigr)
  t^{-m-1}+ \nonumber \\
\!\!\!\!&+\!\!\!\!&\Bigl((\mu^2+\frac{2}{3}\Lambda)\phi_{-m}-\frac{3}{2}m
  \bigl(2\frac{S_{n+2}}{S_n}-\frac{S^2_{n+1}}{S^2_n}\bigr)\phi_{-m}-
  \frac{3}{2}(m-1)\frac{S_{n+1}}{S_n}\phi_{-m+1}+ \nonumber \\
\!\!\!\!&{}\!\!\!\!&+(m-2)(m-1-\frac{3}{2}n)\phi_{-m+2}\Bigr)t^{-m}+ 
  \nonumber \\
\!\!\!\!&+\!\!\!\!&\Bigl(-\frac{3}{2}m\bigl(3\frac{S_{n+3}}{S_n}-3\frac{S_{n+1}
  S_{n+2}}{S^2_n}+\frac{S^3_{n+1}}{S^3_n}\bigr)\phi_{-m}+(\mu^2+\frac{2}{3}
  \Lambda)\phi_{-m+1}-  \nonumber \\
\!\!\!\!&{}\!\!\!\!&-\frac{3}{2}(m-1)\bigl(2\frac{S_{n+2}}{S_n}-\frac{S^2_{n+1}}
  {S^2_n}\bigr)\phi_{-m+1}-\frac{3}{2}(m-2)\frac{S_{n+1}}{S_n}\phi_{-m+2}+
\nonumber \\
\!\!\!\!&{}\!\!\!\!&+(m-3)(m-2-\frac{3}{2}n)\phi_{-m+3}\Bigr)t^{-m+1}+ 
  {\cal O}(t^{-m+2})+ \nonumber \\
\!\!\!\!&+\!\!\!\!&(\mu^2+6\frac{\lambda}{\kappa})t^{-3m}\Bigl(\phi_{-m}+
  \phi_{-m+1}t+\phi_{-m+2}t^2+\phi_{-m+3}t^3+{\cal O}(t^4)\Bigr)^3. \label{eq:3.2}
\end{eqnarray}
The sum of the two Einstein equations in (\ref{eq:1.3}) is 

\begin{eqnarray}
0\!\!\!\!&=\!\!\!\!&\frac{n(n-1)}{2}\frac{1}{t^2}+n\frac{S_{n+1}}{S_n}
  \frac{1}{t}+\Bigl((1+2n)\frac{S_{n+2}}{S_n}-n\frac{S^2_{n+1}}{S^2_n}-
  \frac{2}{3}\Lambda\Bigr)+ \nonumber \\
\!\!\!\!&+\!\!\!\!&\Bigl(3(n+1)\frac{S_{n+3}}{S_n}-(3n+1)\frac{S_{n+1}
  S_{n+2}}{S^2_n}+n\frac{S^3_{n+1}}{S^3_n}\Bigr)t+{\cal O}(t^2)+ \nonumber \\
\!\!\!\!&+\!\!\!\!&\frac{k}{t^n}\Bigl(\frac{1}{S_n}-\frac{S_{n+1}}{S^2_n}t+
  \bigl(\frac{S^2_{n+1}}{S^3_n}-\frac{S_{n+2}}{S^2_n}\bigr)t^2-\bigl(
  \frac{S_{n+3}}{S^2_n}-2\frac{S_{n+1}S_{n+2}}{S^3_n}+\frac{S^3_{n+1}}{S^4_n}
  \bigr)t^3+{\cal O}(t^4)\Bigr)- \nonumber \\
\!\!\!\!&-\!\!\!\!&\frac{\mu^2}{t^{2m}}\Bigl(\phi^2_{-m}+2\phi_{-m}\phi_{-m+1}t+
  \bigl(2\phi_{-m}\phi_{-m+2}+\phi^2_{-m+1}\bigr)t^2+\nonumber \\
\!\!\!\!&{}\!\!\!\!&+2\bigl(\phi_{-m}\phi_{-m+3}+\phi_{-m+1}\phi_{-m+2}\bigr)t^3
  +{\cal O}(t^4)\Bigr); \label{eq:3.3}
\end{eqnarray}
while their difference is 

\begin{eqnarray}
0\!\!\!\!&=\!\!\!\!&\frac{4\kappa}{3}\varepsilon-\frac{n}{t^2}+\Bigl(2
  \frac{S_{n+2}}{S_n}-\frac{S^2_{n+1}}{S^2_n}\Bigr)+2\Bigl(3\frac{S_{n+3}}{S_n}
  -3\frac{S_{n+1}S_{n+2}}{S^2_n}+\frac{S^3_{n+1}}{S^3_n}\Bigr)t+{\cal O}(t^2)
  + \nonumber \\
\!\!\!\!&-\!\!\!\!&\frac{2k}{t^n}\Bigl(\frac{1}{S_n}-\frac{S_{n+1}}{S^2_n}t+
  \bigl(\frac{S^2_{n+1}}{S^3_n}-\frac{S_{n+2}}{S^2_n}\bigr)t^2-\bigl(
  \frac{S_{n+3}}{S^2_n}-2\frac{S_{n+1}S_{n+2}}{S^3_n}+\frac{S^3_{n+1}}{S^4_n}
  \bigr)t^3+{\cal O}(t^4)\Bigr)- \nonumber \\
\!\!\!\!&-\!\!\!\!&\frac{2\mu^2}{t^{2m}}\Bigl(\phi^2_{-m}+2\phi_{-m}\phi_{-m+1}t+
  \bigl(2\phi_{-m}\phi_{-m+2}+\phi^2_{-m+1}\bigr)t^2+\nonumber \\
\!\!\!\!&{}\!\!\!\!&+2\bigl(\phi_{-m}\phi_{-m+3}+\phi_{-m+1}\phi_{-m+2}\bigr)t^3
  +{\cal O}(t^4)\Bigr). \label{eq:3.4}
\end{eqnarray}
We specify the energy density (\ref{eq:1.1}) in the next subsection.

\subsection{The asymptotic solutions with a Big Bang singularity}
\label{sub-3.2}

If $m$ in (\ref{eq:3.2}) were greater than $1$, then $3m>m+2$ would hold, 
and hence the leading order term would be $(\mu^2+6\frac{\lambda}{\kappa})
\phi^3_{-m}t^{-3m}$. However, its vanishing would yield $\phi_{-m}=0$, which 
is a contradiction. Similarly, if $1>m>1/2$ or $1/2>m>0$ held, then the 
power $-3m$ would differ from $-m-2$, $-m-1$, $-m$ and $-m+1$. Hence the 
only term of order $t^{-3m}$ would be $(\mu^2+6\frac{\lambda}{\kappa})\phi
^3_{-m}t^{-3m}$ again, implying the contradiction $\phi_{-m}=0$. Therefore, 
$m$ could be only $1$ or $1/2$. 

Thus first suppose that $m=1$. Then, in the leading order, equations 
(\ref{eq:3.2})-(\ref{eq:3.4}) take the form 

\begin{eqnarray}
&{}&0=\Bigl((\mu^2+6\frac{\lambda}{\kappa})\phi^2_{-1}+2-\frac{3}{2}n\Bigr)
  \phi_{-1}\frac{1}{t^3}+{\cal O}(t^{-2}), \label{eq:3.5a} \\
&{}&0=\Bigl(-2\mu^2\phi^2_{-1}+n(n-1)\Bigr)\frac{1}{t^2}+\frac{2k}{S_n}\frac{1}
  {t^n}+{\cal O}(t^{-1})+{\cal O}(t^{-n+1}), \label{eq:3.5b} \\
&{}&0=\Bigl(\bigl(\mu^2+6\frac{\lambda}{\kappa}\bigr)\phi^2_{-1}+2-\frac{3}{2}n
  \Bigr)\frac{1}{t^2}+\frac{k}{S_n}\frac{1}{t^n}+{\cal O}(t^{-1})+{\cal O}
  (t^{-n+1}), \label{eq:3.5c}
\end{eqnarray}
where, to derive (\ref{eq:3.5c}), we used the leading order expression 

\begin{equation*}
\frac{4\kappa}{3}\varepsilon=-4\Bigl(3\frac{\lambda}{\kappa}\phi^2_{-1}+1-n
\Bigr)\frac{1}{t^2}+{\cal O}(t^{-1})
\end{equation*}
of the energy density (\ref{eq:1.3}). We show that equations 
(\ref{eq:3.5a})-(\ref{eq:3.5c}) lead to contradictions, i.e. the field 
equations do not have any asymptotic power series solution with $m=1$. 

If $n>2$ held, then the leading order term in (\ref{eq:3.5b}) would be 
$(k/S_n)t^{-n}$, whose vanishing would yield $k=0$. But then the leading order 
term in (\ref{eq:3.5b}) with $k=0$ would be of order $t^{-2}$, the vanishing 
of whose coefficient would give $n(n-1)=2\mu^2\phi^2_{-1}$. Since $\mu^2<0$, 
this is a contradiction. Similarly, if $2>n>1$, then the only term of order 
$t^{-n}$ in (\ref{eq:3.5b}) would be $(k/S_n)t^{-n}$, whose vanishing would 
give $k=0$, which would leave us to the contradiction above again. If $n=2$, 
then from (\ref{eq:3.5a}) and (\ref{eq:3.5c}) it follows that $k=0$, which, 
by (\ref{eq:3.5b}), implies that $\mu^2\phi^2_{-1}=1$. However, by $\mu^2<0$ 
this is a contradiction. If $n=1$, then the leading order term in 
(\ref{eq:3.5b}) is $-\mu^2\phi^2_{-1}t^{-2}$, whose vanishing yields the 
contradiction $\phi_{-1}=0$. Finally, if $1>n>0$, then (\ref{eq:3.5a}) gives 
$2(\mu^2+6\lambda/\kappa)\phi^2_{-1}=3n-4$, whose right hand side is negative, 
while $\mu^2+6\lambda/\kappa$ is positive, which is a contradiction. 

Next suppose that $m=1/2$ in the expansion (\ref{eq:3.1a}). Then the leading 
order term in (\ref{eq:3.2}) is $\frac{3}{4}(1-n)\phi_{-1/2}t^{-5/2}$, whose 
vanishing implies $n=1$. With this substitution equations (\ref{eq:3.2}) and 
(\ref{eq:3.3}), respectively, give 

\begin{eqnarray}
&{}&2\phi_{\frac{1}{2}}=3\frac{S_2}{S_1}\phi_{-\frac{1}{2}}-4\bigl(\mu^2+6
  \frac{\lambda}{\kappa}\bigr)\phi^3_{-\frac{1}{2}}, \label{eq:3.6a} \\
&{}&3\phi_{\frac{3}{2}}=\frac{3}{2}\frac{S_3}{S_1}\phi_{-\frac{1}{2}}-
  \frac{3}{4}\frac{S_2}{S_1}\bigl(\frac{S_2}{S_1}\phi_{-\frac{1}{2}}+\phi
  _{\frac{1}{2}}\bigr)-\bigl(\mu^2+\frac{2}{3}\Lambda\bigr)\phi_{-\frac{1}{2}}
  -3\bigl(\mu^2+6\frac{\lambda}{\kappa}\bigr)\phi^2_{-\frac{1}{2}}\phi
  _{\frac{1}{2}}, \label{eq:3.6b}\\
&{}&\frac{15}{2}\phi_{\frac{5}{2}}=\frac{9}{4}\frac{S_4}{S_1}\phi_{-\frac{1}{2}}-
  \frac{3}{2}\frac{S_3}{S_1}\phi_{\frac{1}{2}}-\frac{9}{4}\frac{S_3S_2}{S^2_1}
  \phi_{-\frac{1}{2}}+\frac{3}{4}\frac{S^2_2}{S^2_1}\phi_{\frac{1}{2}}+\frac{3}{4}
  \frac{S^3_2}{S^3_1}\phi_{-\frac{1}{2}}-\frac{9}{4}\frac{S_2}{S_1}\phi_{\frac{3}{2}}
  - \nonumber \\
&{}&\hskip 30pt-\bigl(\mu^2+\frac{2}{3}\Lambda\bigr)\phi_{\frac{1}{2}}-3\bigl(
  \mu^2+6\frac{\lambda}{\kappa}\bigr)\phi_{-\frac{1}{2}}\bigl(\phi_{\frac{3}{2}}
  \phi_{-\frac{1}{2}}+\phi^2_{\frac{1}{2}}\bigr); \label{eq:3.6c}
\end{eqnarray}
and 

\begin{eqnarray}
&{}&\frac{S_2}{S_1}=-\frac{k}{S_1}+\mu^2\phi^2_{-\frac{1}{2}}, \label{eq:3.7a} \\
&{}&3\frac{S_3}{S_1}=\frac{S^2_2}{S^2_1}+k\frac{S_2}{S^2_1}+\frac{2}{3}\Lambda
  +2\mu^2\phi_{-\frac{1}{2}}\phi_{\frac{1}{2}}, \label{eq:3.7b}\\
&{}&6\frac{S_4}{S_1}=4\frac{S_2S_3}{S^2_1}-\frac{S^3_2}{S^3_1}+k\frac{S_3}
  {S^2_1}-k\frac{S^2_2}{S^3_1}+\mu^2\bigl(2\phi_{\frac{3}{2}}\phi_{-\frac{1}{2}}
  +\phi^2_{\frac{1}{2}}\bigr). \label{eq:3.7c}
\end{eqnarray}
These equations form a hierarchical system, and can be solved successively 
for $S_2$, $\phi_{\frac{1}{2}}$, $S_3$, $\phi_{\frac{3}{2}}$, $S_4$ and $\phi
_{\frac{5}{2}}$ in terms of $\phi_{-\frac{1}{2}}$, $S_1>0$ and the discrete 
parameter $k$. 

The ${\cal O}(t)$ accurate expression of the energy density is 

\begin{eqnarray*}
\frac{4\kappa}{3}\varepsilon\!\!\!\!&=\!\!\!\!&\frac{1}{t^2}+\Bigl(\frac{1}
  {\phi^2_{-\frac{1}{2}}}+2\frac{S_2}{S_1}-12\frac{\lambda}{\kappa}\phi^2
  _{-\frac{1}{2}}\Bigr)\frac{1}{t}+\Bigl(4\frac{S_3}{S_1}-2\frac{S^2_2}{S^2_1}-
  4\frac{\phi_{\frac{1}{2}}}{\phi_{-\frac{1}{2}}}\frac{S_2}{S_1}+\frac{2}{\phi^2
  _{-\frac{1}{2}}}\frac{S_2}{S_1}- \\
\!\!\!\!&{}\!\!\!\!&-4\frac{\phi^2_{\frac{1}{2}}}{\phi^2_{-\frac{1}{2}}}-2
  \frac{\phi_{\frac{1}{2}}}{\phi^3_{-\frac{1}{2}}}+\frac{1}{\phi^4_{-\frac{1}{2}}}-24
  \frac{\lambda}{\kappa}\phi_{\frac{1}{2}}\phi_{-\frac{1}{2}}-4\bigl(\mu^2+
  \frac{\Lambda}{3}+3\frac{\lambda}{\kappa}\bigr)\Bigr)+ \\
\!\!\!\!&+\!\!\!\!&\Bigl(6\frac{S_4}{S_1}-6\frac{S_2S_3}{S^2_1}+2\frac{S^3_2}
  {S^3_1}+\frac{4}{\phi^2_{-\frac{1}{2}}}\frac{S_3}{S_1}(1-2\phi_{\frac{1}{2}}\phi
  _{-\frac{1}{2}})-\frac{2}{\phi^2_{-\frac{1}{2}}}\frac{S^2_2}{S^2_1}(1-2
  \phi_{\frac{1}{2}}\phi_{-\frac{1}{2}})+ \\
\!\!\!\!&{}\!\!\!\!&+2\frac{S_2}{S_1}\bigl(-4\frac{\phi_{\frac{3}{2}}}{\phi
  _{-\frac{1}{2}}}+2\frac{\phi^2_{\frac{1}{2}}}{\phi^2_{-\frac{1}{2}}}-4\frac{\phi
  _{\frac{1}{2}}}{\phi^3_{-\frac{1}{2}}}+\frac{1}{\phi^4_{-\frac{1}{2}}}\bigr)-16
  \frac{\phi_{\frac{3}{2}}\phi_{\frac{1}{2}}}{\phi^2_{-\frac{1}{2}}}-2\frac{\phi
  _{\frac{3}{2}}}{\phi^3_{-\frac{1}{2}}}+8\frac{\phi^3_{\frac{1}{2}}}{\phi^3
  _{-\frac{1}{2}}}- \\
\!\!\!\!&{}\!\!\!\!&-\frac{\phi^2_{\frac{1}{2}}}{\phi^4_{-\frac{1}{2}}}-4\frac{\phi
  _{\frac{1}{2}}}{\phi^5_{-\frac{1}{2}}}+\frac{1}{\phi^6_{-\frac{1}{2}}}-\frac{4}{\phi
  ^2_{-\frac{1}{2}}}\bigl(\mu^2+\frac{\Lambda}{3}+3\frac{\lambda}{\kappa}\bigr)-
  12\frac{\lambda}{\kappa}\bigl(\phi^2_{\frac{1}{2}}+2\phi_{\frac{3}{2}}\phi
  _{-\frac{1}{2}}\bigr)\Bigr)t+{\cal O}(t^2). 
\end{eqnarray*}
With this substitution the field equation (\ref{eq:3.4}) (with $m=1/2$ and 
$n=1$) is identically satisfied in the ${\cal O}(t^{-2})$ order, and the 
requirement of the vanishing of its ${\cal O}(t^{-1})$ order part yields 

\begin{equation}
2\frac{S_2}{S_1}=-\frac{1}{\phi^2_{-\frac{1}{2}}}+2\frac{k}{S_1}+2\mu^2\phi^2
_{-\frac{1}{2}}+12\frac{\lambda}{\kappa}\phi^2_{-\frac{1}{2}}. \label{eq:3.8a}
\end{equation}
Comparing (\ref{eq:3.8a}) with (\ref{eq:3.7a}), we find that 

\begin{equation}
\frac{4k}{S_1}=\frac{1}{\phi^2_{-\frac{1}{2}}}-12\frac{\lambda}{\kappa}\phi^2
_{-\frac{1}{2}}. \label{eq:3.9}
\end{equation}
Thus, if $k=0$, then $\phi^4_{-\frac{1}{2}}=\kappa/12\lambda$, i.e. it is 
already fixed by the parameters of the model, but $S_1$ is still arbitrary, 
but positive. If $k=\pm1$, then $S_1$ is determined by $\phi_{-\frac{1}{2}}$ by 
(\ref{eq:3.9}). But since $S_1$ must be positive, the domain of $\phi^2
_{-\frac{1}{2}}$ is restricted: $\phi^4_{-\frac{1}{2}}<\kappa/12\lambda$ if $k=1$, 
and $\phi^4_{-\frac{1}{2}}>\kappa/12\lambda$ if $k=-1$. 

The structure of the formulae (\ref{eq:3.6a})-(\ref{eq:3.7c}) shows that 
when we intend to express $S_2$, $\phi_{\frac{1}{2}}$, $S_3$, $\phi_{\frac{3}{2}}$, 
$S_4$ and $\phi_{\frac{5}{2}}$ in terms of the remaining free parameter (i.e. 
by $\phi_{-\frac{1}{2}}$ if $k=\pm1$ and by $S_1$ if $k=0$) via (\ref{eq:3.9}), 
the resulting formulae \emph{will be expressions of $\phi_{-\frac{1}{2}}$ alone 
even when $k=0$}. Thus, $S_1$ plays the role only of an uninteresting uniform 
scale factor that fixes the unit of spatial length. 

In the next two orders, the field equation (\ref{eq:3.4}) yields 

\begin{eqnarray}
6\frac{S_3}{S_1}\!\!\!\!&=\!\!\!\!&3\frac{S^2_2}{S^2_1}+4\frac{\phi_{\frac{1}{2}}}
  {\phi_{-\frac{1}{2}}}\frac{S_2}{S_1}-\frac{2}{\phi^2_{-\frac{1}{2}}}\frac{S_2}
  {S_1}-2k\frac{S_2}{S^2_1}-\frac{1}{\phi^4_{-\frac{1}{2}}}+4\frac{\phi^2
  _{\frac{1}{2}}}{\phi^2_{-\frac{1}{2}}}+2\frac{\phi_{\frac{1}{2}}}{\phi^3
  _{-\frac{1}{2}}}- \nonumber \\
\!\!\!\!&{}\!\!\!\!& -4\bigl(\mu^2+6\frac{\lambda}{\kappa}\bigr)
  \phi_{\frac{1}{2}}\phi_{-\frac{1}{2}}+4\bigl(\mu^2+\frac{\Lambda}{3}+3
  \frac{\lambda}{\kappa}\bigr), \label{eq:3.8b}\\
12\frac{S_4}{S_1}\!\!\!\!&=\!\!\!\!&12\frac{S_2S_3}{S^2_1}-4\frac{S^3_2}{S^3_1}
  +\frac{S^2_2}{S^2_1}\frac{2}{\phi^2_{-\frac{1}{2}}}\bigl(1+\frac{k}{S_1}\phi^2
  _{-\frac{1}{2}}-2\phi_{\frac{1}{2}}\phi_{-\frac{1}{2}}\bigr)- \nonumber \\
\!\!\!\!&{}\!\!\!\!&-\frac{S_3}{S_1}\frac{2}{\phi^2_{-\frac{1}{2}}}\bigl(2+
  \frac{k}{S_1}\phi^2_{-\frac{1}{2}}-4\phi_{\frac{1}{2}}\phi_{-\frac{1}{2}}\bigr)+2
  (\mu^2+6\frac{\lambda}{\kappa})\bigl(\phi^2_{\frac{1}{2}}+2\phi_{\frac{3}{2}}\phi
  _{-\frac{1}{2}}\bigr)+\nonumber \\
\!\!\!\!&{}\!\!\!\!&+\frac{4}{\phi^2_{-\frac{1}{2}}}\bigl(\mu^2+\frac{\Lambda}{3}
  +3\frac{\lambda}{\kappa}\bigr)+2\frac{S_2}{S_1}\bigl(4\frac{\phi_{\frac{3}{2}}}
  {\phi_{-\frac{1}{2}}}-2\frac{\phi^2_{\frac{1}{2}}}{\phi^2_{-\frac{1}{2}}}+4
  \frac{\phi_{\frac{1}{2}}}{\phi^3_{-\frac{1}{2}}}-\frac{1}{\phi^4_{-\frac{1}{2}}}
  \bigr)+ \nonumber \\
\!\!\!\!&{}\!\!\!\!&+16\frac{\phi_{\frac{3}{2}}\phi_{\frac{1}{2}}}{\phi^2
  _{-\frac{1}{2}}}+2\frac{\phi_{\frac{3}{2}}}{\phi^3_{-\frac{1}{2}}}-8\frac{\phi^3
  _{\frac{1}{2}}}{\phi^3_{-\frac{1}{2}}}+\frac{\phi^2_{\frac{1}{2}}}{\phi^4
  _{-\frac{1}{2}}}+4\frac{\phi_{\frac{1}{2}}}{\phi^5_{-\frac{1}{2}}}-\frac{1}{\phi^6
  _{-\frac{1}{2}}}. \label{eq:3.8c}
\end{eqnarray}
Then, it is a lengthy but straightforward calculation to check that, as a 
consequence of (\ref{eq:3.6a})-(\ref{eq:3.7c}) and (\ref{eq:3.9}), these 
equations are identically satisfied. 

Therefore, (\ref{eq:3.1a})-(\ref{eq:3.1b}) with $n=1$ and $m=1/2$ and 
expansion coefficients satisfying (\ref{eq:3.6a})-(\ref{eq:3.7c}) and 
(\ref{eq:3.9}) provide an asymptotic solution of the field equations. For 
$k=0$ this solution is uniquely determined (up to an overall scale parameter), 
but for $k=\pm1$ this is in fact a whole 1-parameter family of solutions. 
Since $R=4\Lambda+\kappa\mu^2\Phi^2\sim-1/t$ and $R_{ab}R^{ab}\sim1/t^4$, the 
instant $t=0$ corresponds to a physical, scalar polynomial curvature 
singularity of the spacetime (Big Bang). Since in a neighbourhood of the 
singularity the mean curvature diverges, in a neighbourhood of the Big Bang 
it plays the role of a correct time function, and also the rest mass of the 
Higgs field cannot be defined and the BEH mechanism does not work.


\section{Numerical results}
\label{sec:4}

To clarify the global properties and the large scale behaviour of the 
solutions of the \emph{exact} field equations (\ref{eq:1.1})-(\ref{eq:1.2}) 
of the EccH system, global techniques or numerical calculations should be 
used. These could provide links between the asymptotic solutions obtained 
in the different situations. Also, new, unexpected qualitative behaviour of 
them may be revealed. 

In this section we present the results of numerical calculations. The role 
of the asymptotic solutions above in these calculations is to provide only 
the appropriate \emph{initial conditions} in physically well formulated 
situations. 

In the numerical calculations near the initial singularities it is natural 
to use the Planck units. In Planck length units, $L_P:=\sqrt{\hbar G/c^3}=
1.6\times10^{-33}cm$, the dimensional parameters of the EccH model are $\mu^2
=-4.6\times10^{-35}L^{-2}_P$, $\Lambda=2.6\times10^{-124}L^{-2}_P$, $6/\kappa=
2.2\times10^{-1}L^{-2}_P$ and $\frac{1}{9}\chi^2_c=1.4\times10^{-2}L^{-2}_P$.  
Hence, in these calculations, the cosmological constant can be considered to 
be zero, and the rest-mass parameter $\mu^2$ of the Higgs field plays only a 
marginal role. To obtain proper initial conditions from the expansions, the 
series in equations (\ref{eq:2.1a}) and (\ref{eq:2.1b}) should converge, i.e. 
$\xi t\ll 1$, where $\xi$ is the largest dimensional parameter of the model 
in inverse time units. Since $\sqrt{\kappa}$ is of order $1/T_P$, where $T_P
\simeq5.4\times10^{-44}sec$ is the Planck time, the numerical calculations 
should be started at $t_0\ll T_P$. Hence, in the calculations of the singular 
solutions, we set our initial conditions at $t_0= 0.001 \, T_P$. 

The dynamics of the system is given by the second equation of (\ref{eq:1.1}) 
and equation (\ref{eq:1.2}). The constraint equation, the first of 
(\ref{eq:1.1}), is used for measuring the numerical precision of the 
solutions. For solving this system of ordinary differential equations we 
use the 4th order Runge-Kutta method with adaptive stepsize control 
\cite{num-rec}. 

All of the numerical solutions fulfill the constraint equation with very high 
precision: the difference of the two sides of the constraint is always less 
than $10^{-4}L^{-2}_P$. In all the figures the numerical inaccuracy is smaller 
than the line widths used in the figures. With the numerical method we use 
here we could reach approximately $10^{22} T_P\simeq10^{-22}\,sec$ (while the 
characteristic time scale of the weak interactions is only $\sim10^{-27}\,sec$).

\subsection{Solutions with a Small Bang singularity}
\label{sub-4.1}

\begin{figure}[ht]
\begin{center}
\includegraphics[width=0.45\textwidth]{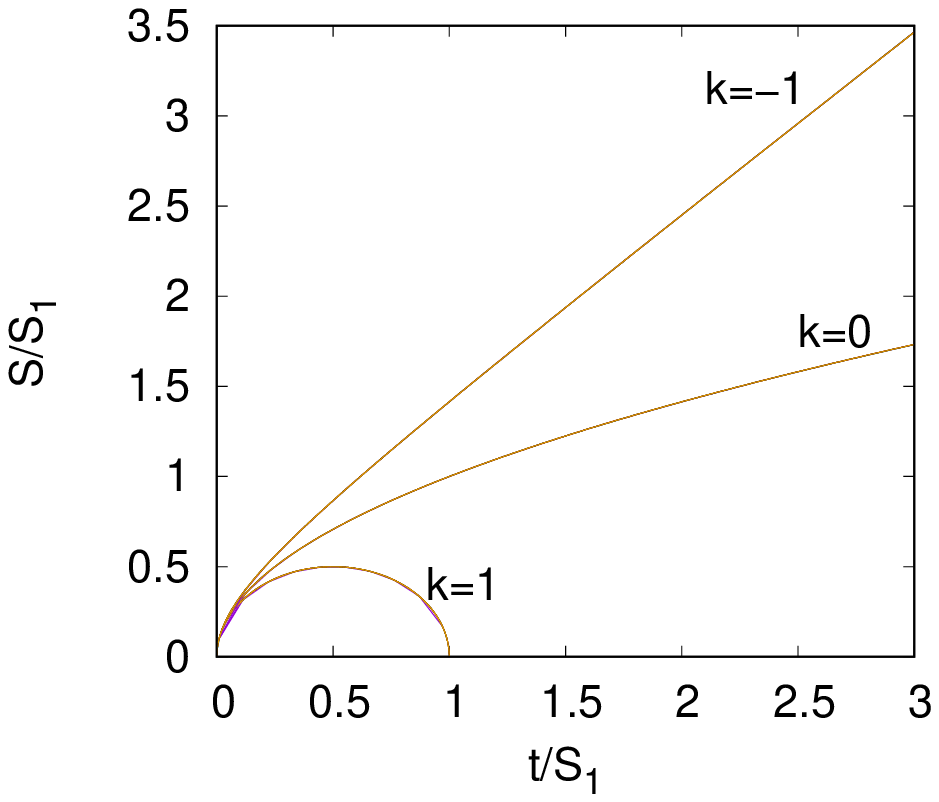}
\includegraphics[width=0.45\textwidth]{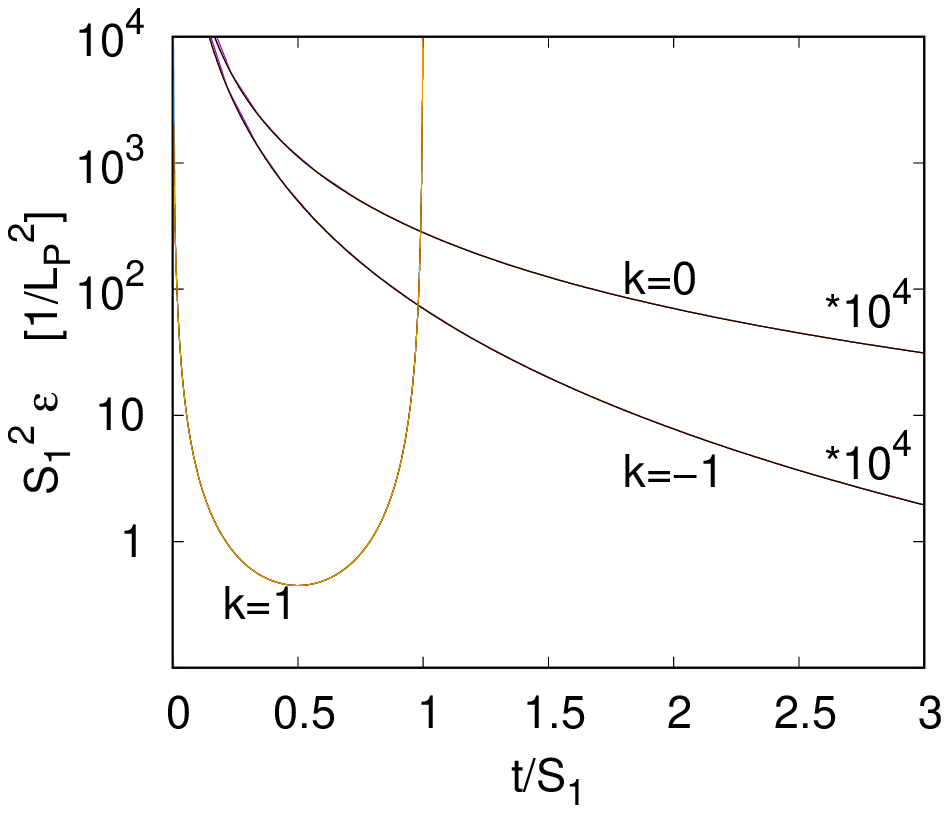}
\caption{\label{fig:SBN1}
The rescaled scale function (left) and energy density (right) in the solutions 
with a Small Bang singularity as a function of the scaled time for $k=0,
\pm1$. In the right figure the $k=0$ and $k=-1$ lines are multiplied by 
$10000$. Each line in the figures is actually six lines on top of each other 
for $S_1$ changing from $10^{-1}$ to $10^4$ for the same $k$. The universe 
recollapses for $k=1$, but expands forever for $k=0,-1$.}
\end{center}
\end{figure}

By (\ref{eq:2.5a})-(\ref{eq:2.5c}) and (\ref{eq:2.7}) the scale function and 
the energy density in the asymptotic solution with the Small Bang singularity 
can be written, respectively, as 

\begin{eqnarray*}
\frac{S^2(t)}{S^2_1}\!\!\!\!&=\!\!\!\!&(\frac{t}{S_1})-k(\frac{t}{S_1})^2+
  \frac{1}{3}(\mu^2+\frac{2}{3}\Lambda)S^2_1(\frac{t}{S_1})^3-\frac{1}{6}
  (\mu^2+\frac{2}{3}\Lambda)kS^2_1(\frac{t}{S_1})^4+{\cal O}(t^5), \\
\kappa\varepsilon(t)S^2_1\!\!\!\!&=\!\!\!\!&\frac{3}{4}(\frac{S_1}{t})^2+
  \frac{3}{2}k(\frac{S_1}{t})+\frac{9}{4}k^2+(\mu^2-\frac{\Lambda}{3})S^2_1
  +{\cal O}(t). 
\end{eqnarray*}
Therefore, in the first two and three terms, respectively, there is a 
scaling property: The scale function, measured in units of the freely 
specifiable initial datum $S_1$ and considered to be a function of $t/S_1$, 
is \emph{independent} of the initial datum; and the same holds for the 
combination $\varepsilon S^2_1$, too. (See Fig.~\ref{fig:SBN1}.) (It might be 
worth noting that $S^2_1$ is just the leading term in $S^4(t)$, which appears 
as a weight function in front of the energy density in the `conservation law' 
$\nabla_aT^a{}_b=0$ mentioned in subsection \ref{sub-1.2}.) Since the order of 
magnitude of the coefficients of the next terms in which the initial datum 
$S_1$ appears in itself is only about $10^{-35}$ times the preceding terms, 
even in the numerical calculations it seems natural to compute $S(t)/S_1$ and 
$\varepsilon S^2_1$ as a function of $t/S_1$. Therefore, the asymptotic 
solution, already with its first few terms, provides a surprisingly good 
approximation of the exact solution: The curves corresponding to different 
initial conditions $S_1$ coincide even though $S_1$ is changing from $10^{-1}$ 
to $10^4$. For $k=0,-1$ the universe is expanding forever, but it recollapses 
for $k=1$.

\begin{figure}[ht]
\begin{center}
\includegraphics[width=0.33\textwidth]{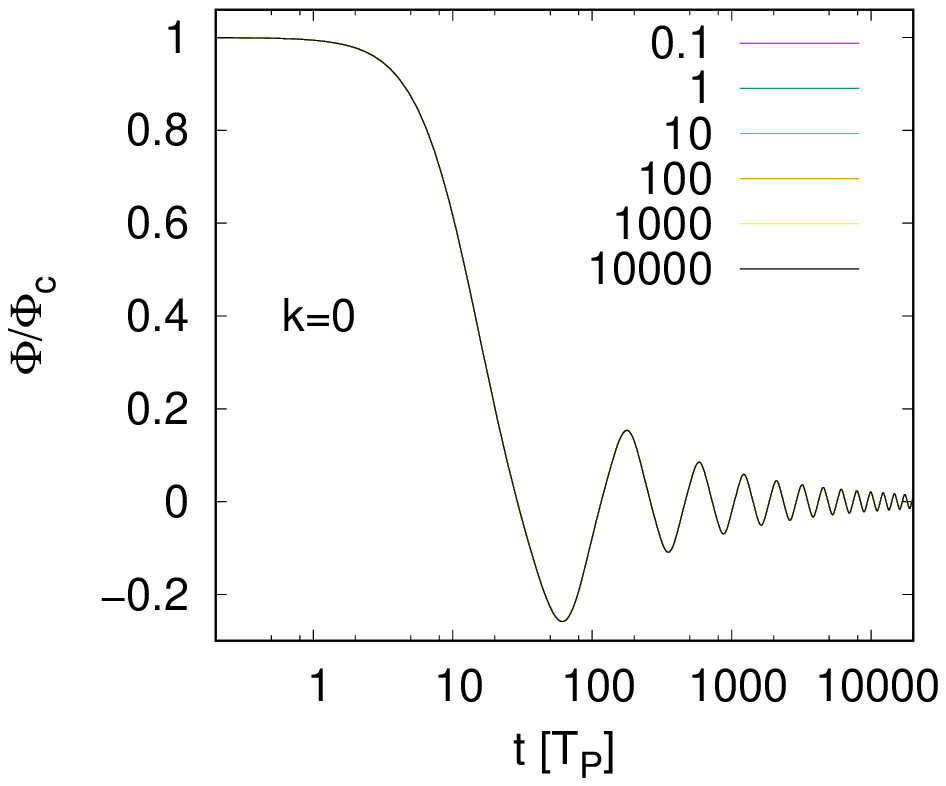}
\includegraphics[width=0.33\textwidth]{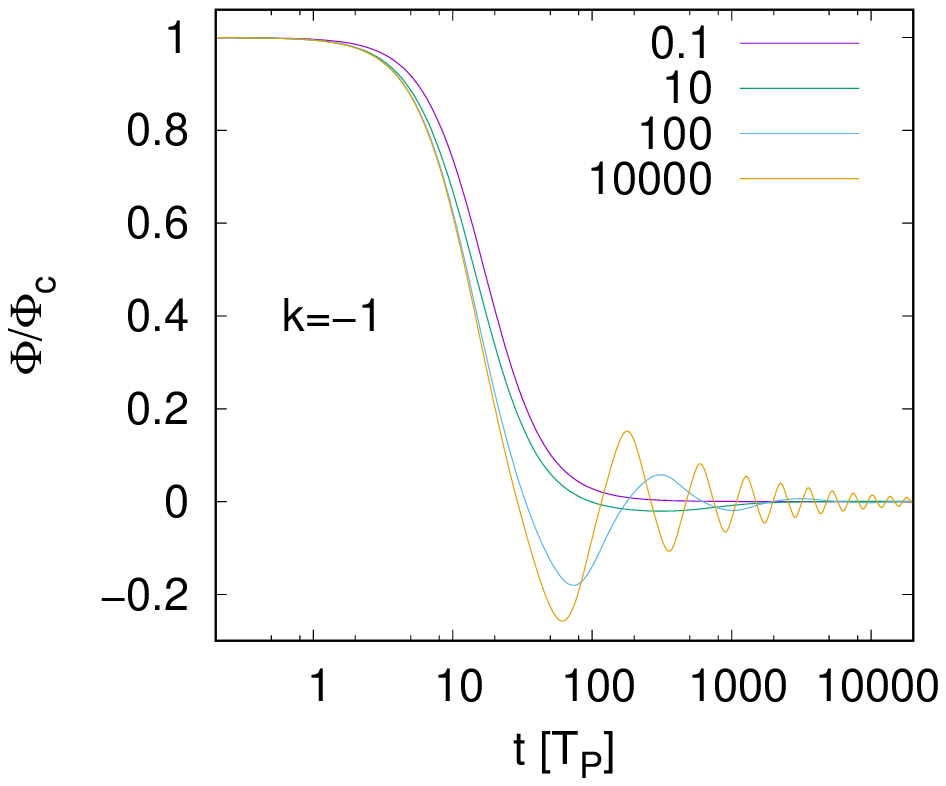}
\includegraphics[width=0.31\textwidth]{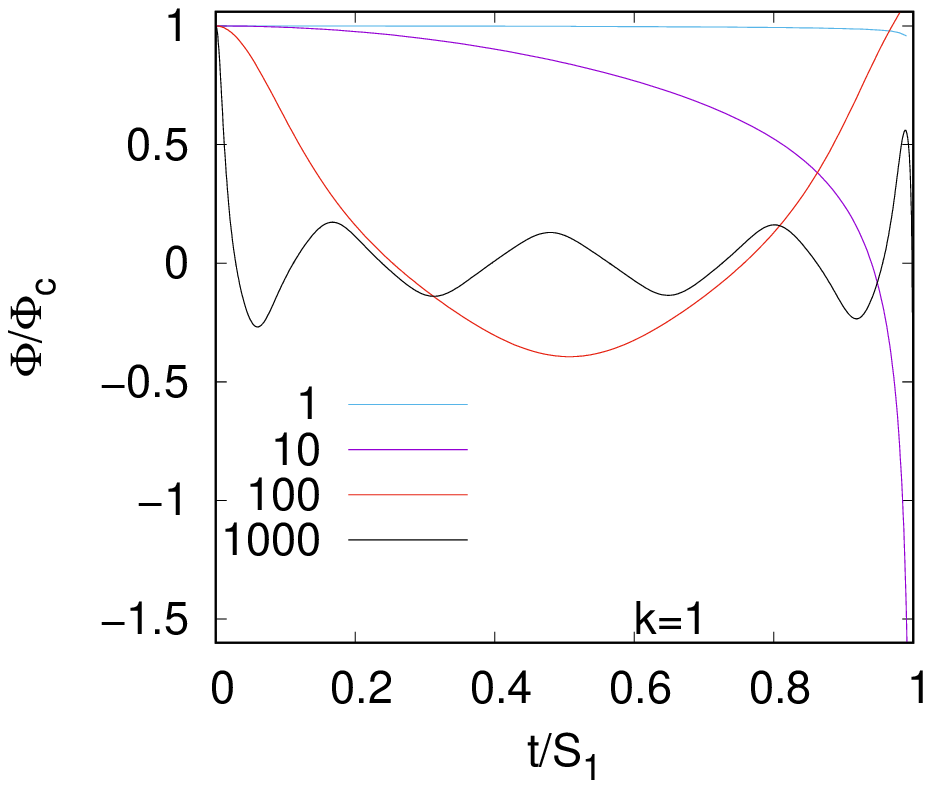}
\caption{\label{fig:SBN1-Phi}
The Higgs field in the solutions with a Small Bang singularity for $k=0,\pm1$ 
and for $S_1$ changing from $10^{-1}$ to $10^4$. In the $k=0$ case $\Phi$ does 
not depend on $S_1$. In the $k=1$ case we use the time coordinate $t/S_1$, in 
which the total `lifetime' of the universe is $1$, independently of $S_1$. 
$\Phi_c$ denotes $\sqrt{6/\kappa}$. }
\end{center}
\end{figure}

The mean curvature in all the three cases $k=0,\pm1$ is strictly monotonically 
decreasing (hence it can indeed be used as a correct time coordinate, the 
`York time'), and it shows similar universal, scaling character. In case of 
$k=0$ even the $S_1$ dependence disappears (see equation (\ref{eq:2.8})). 
The Higgs field also has this scaling property in its first two orders: 

\begin{eqnarray*}
\sqrt{\frac{\kappa}{6}}\Phi(t)-1=S^2_1\Bigl(\!\!\!\!&-\!\!\!\!&\frac{2}{5}
  \bigl(3\frac{\lambda}{\kappa}+\mu^2+\frac{\Lambda}{3}\bigr)(\frac{t}{S_1})
  ^2-\frac{4}{35}\bigl(3\frac{\lambda}{\kappa}+\mu^2+\frac{\Lambda}{3}\bigr)
  k(\frac{t}{S_1})^3+ \\
\!\!\!\!&+\!\!\!\!&\frac{2}{15}\bigl(3\frac{\lambda}{\kappa}+\mu^2+
  \frac{\Lambda}{3}\bigr)\bigl(3\frac{\lambda}{\kappa}+\mu^2+\frac{\Lambda}
  {3}-\frac{5}{7}\frac{k^2}{S^2_1}\bigr)S^2_1(\frac{t}{S_1})^4+{\cal O}(t^5)
  \Bigr). 
\end{eqnarray*}
However, this scaling property is not showing up in the numerical solutions, 
because in the third term (where $S_1$ shows up without the $1/t$) the 
numerical coefficient has the same order of magnitude as the previous two. 
Hence this is not negligible, so already around the Planck time this starts 
to play an important role, spoiling the scaling behaviour. In (\ref{eq:1.2}), 
the equation of motion for the Higgs field, the scale factor appears through 
the term $\dot S/S=\chi/3$. Therefore, since for $k=0$ the mean curvature is 
independent of $S_1$, the time evolution of the Higgs field for $k=0$ is 
universal, and the numerical solutions confirm this (see 
Fig.~\ref{fig:SBN1-Phi}).

\subsection{Solutions with a Milne type singularity}
\label{sub-4.2}

In subsection \ref{sub-2.5} we derived the equations for the one-parameter
family of solutions with the Milne type singularity. In a neighbourhood of 
the singularity $\dot S=1+{\cal O}(t)$. However, the numerical calculations 
show that it does not change significantly even on much larger scales, i.e. 
as far as we can calculate, $S(t)={\cal O}(t)$. If $\phi_0>\phi_0^{crit}$ 
(see equation (\ref{eq:2.20})) the Higgs field is decreasing with time 
near $t=0$, but later it oscillates around zero. If $\phi_0<\phi_0^{crit}$, 
then the time derivative of the Higgs field is positive but negligible. Thus 
the Higgs field is although increasing, practically it remains constant.

\subsection{Solutions with a Big Bang singularity}
\label{sub-4.3}

\begin{figure}[ht]
\begin{center}
\includegraphics[width=0.45\textwidth]{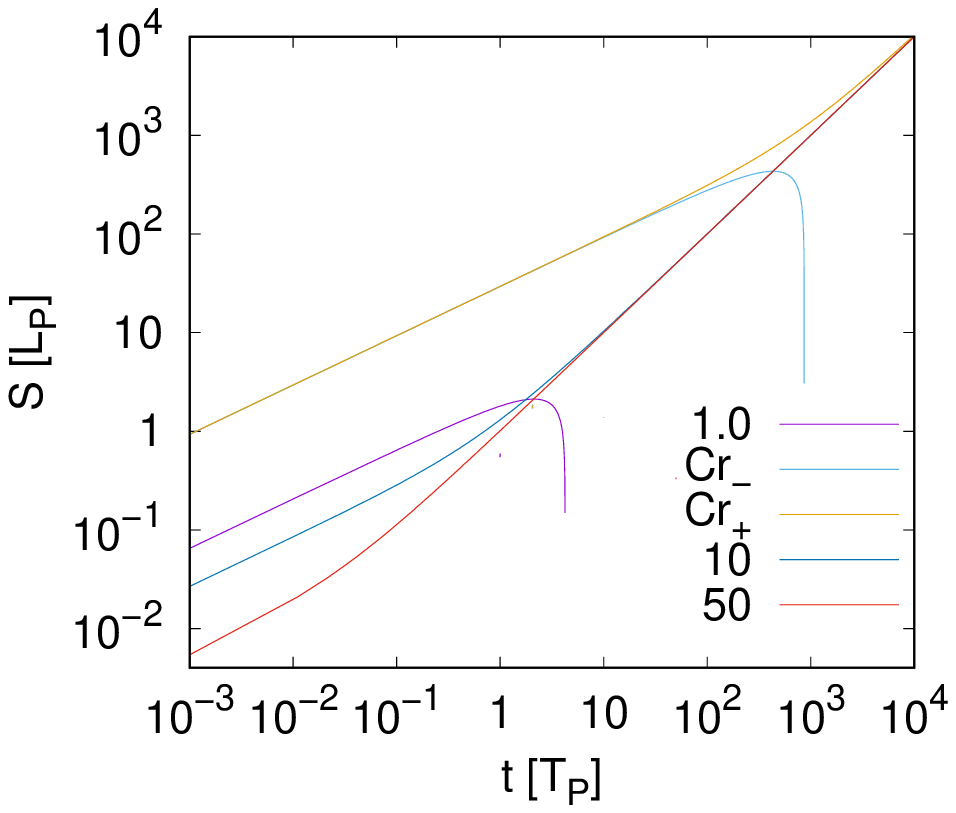}
\includegraphics[width=0.45\textwidth]{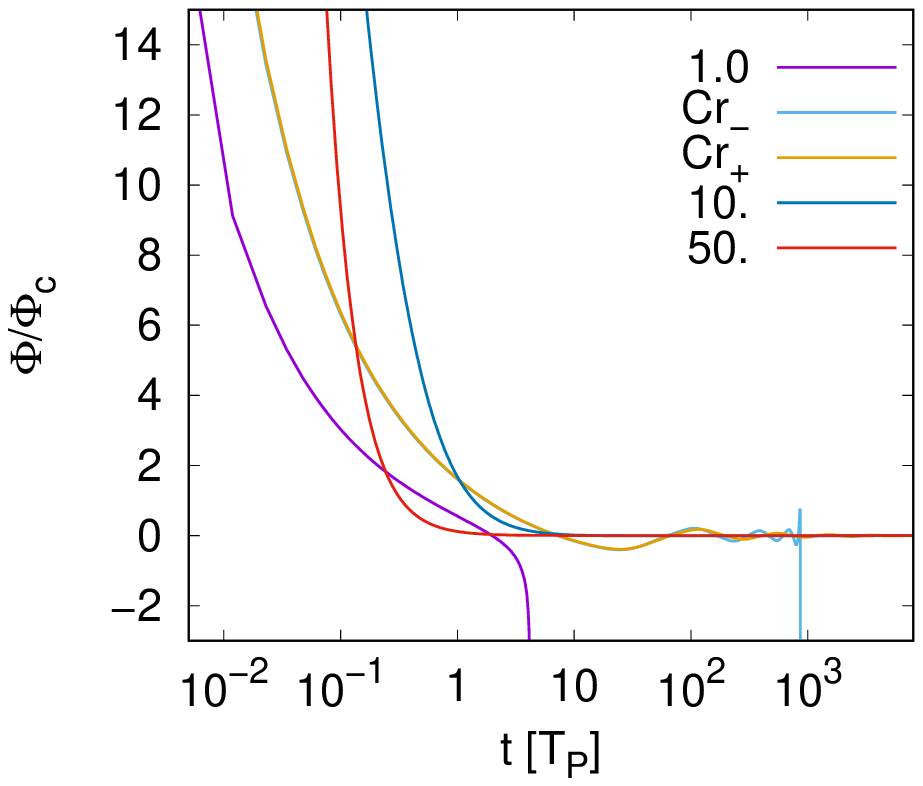}
\caption{\label{fig:BB}
The scale function and the Higgs field as a function of time for various 
values of the parameter $\phi_{-1/2}$ in solutions with a Big Bang singularity. 
The discrete parameter $k$ is necessarily $1$ for $\phi^2_{-1/2}$ less, and it 
is $-1$ for $\phi^2_{-1/2}$ greater than its critical value $Cr:=
\sqrt{\kappa/12\lambda}\simeq2.054312...$; and $k=0$ for the critical value. 
Here $Cr_\pm:=Cr \pm 0.01$. When $k=0$, then the slope of the scale function 
agrees with the slope of the linear part of the scale functions with 
$\phi_{-1/2}$ close to the critical value. $\Phi_c$ denotes $\sqrt{6/\kappa}$.}
\end{center}
\end{figure}

In subsection \ref{sub-3.2} we saw that the asymptotic power series 
solutions with the Big Bang singularity form a 1-parameter family; and for 
the parameter value $\phi^2_{-1/2}$ smaller than $\sqrt{\kappa/12\lambda}$ the 
discrete parameter $k$ is necessarily 1, for $\phi^2_{-1/2}=\sqrt{\kappa/12
\lambda}$ it is zero, while for $\phi^2_{-1/2}>\sqrt{\kappa/12\lambda}$ it is 
$-1$. The universe is necessarily recollapsing for $k=1$, in which case the 
Higgs field diverges (tending either to $\infty$ or $-\infty$) in the `Big 
Crunch' singularity. The moment of the `Big Crunch' depends on the parameter 
$\phi^2_{-1/2}$ of the solution: The closer the parameter $\phi^2_{-1/2}$ to the 
critical value is, the later the time of the `Big Crunch' is (see 
Fig.~\ref{fig:BB}). 

If $k=0$ or $-1$, then the universe is expanding forever, and the rate of 
expansion for large proper time is getting to be independent of the parameter 
of the solution. This behaviour is, indeed, compatible with the asymptotic 
vanishing of the Higgs field.

\subsection{Solutions with initial data on the $\Phi=\sqrt{6/
\kappa}$ hypersurface}
\label{sub-4.4}

\subsubsection{The strategy of the calculations}
\label{sub-4.4.1}

In subsection \ref{sub-2.4} we saw that solutions that are regular near the 
spacelike hypersurface $t=0$ on which $\Phi(0)=\sqrt{6/\kappa}$ holds can be 
parametrized by two continuous parameters, $S_0$ and $S_1$, fulfilling the 
inequalities (\ref{eq:reg}), and the discrete parameter $k$ and the sign of 
$\Pi(0)$, denoted henceforth by $s$. An interesting question is what kind of 
\emph{global} solutions can we have with such initial conditions. On the 
other hand, in the numerical calculations of subsections 
\ref{sub-4.1}--\ref{sub-4.3}, we followed just the opposite strategy: 
Starting from the singularity (at $t=0$) of a singular solution we calculated 
the various quantities at later times $t>0$. 

Clearly, the latter strategy can be used to determine the regular initial 
data for a given solution on a spacelike hypersurface \emph{defined} e.g. by 
some specific value of the Higgs field, $\Phi=\Phi_0$ for some $\Phi_0\geq0$. 
(By the gauge freedom $\mathbb{Z}_2:\Phi\mapsto-\Phi$ one can always assume 
that $\Phi_0\geq0$.) In particular, if $(S(t),\Phi(t))$ is a solution of the 
field equations with an initial singularity at $t=0$ and there is some 
instant when $\Phi=\Phi_0$, then primarily we are interested in the initial 
data for the solution $(S(t),\Phi(t))$ at the instant when $\Phi$ takes the 
value $\Phi_0$ \emph{at first time}. Thus, in the present numerical 
calculations, we use the Higgs field to be the time variable, rather than 
the mean curvature. (This choice for the time variable was motivated by the 
investigations of subsection \ref{sub-2.4}.) Clearly, at this instant $\dot
\Phi\leq0$ holds for solutions developing from a Big Bang type singularity, 
but in general $\dot S>0$ does not necessarily hold. The latter would mean 
that the universe is still \emph{expanding} at the instant when $\Phi=\Phi
_0$. Of course, the set of all the initial data obtained in this way is 
\emph{not} quite the set of all the independent data sets in the usual Cauchy 
problem for the system (or rather the set of the solutions of the field 
equations): E.g. the Cauchy data sets for those solutions of the field 
equations in which $\Phi$ remains less than $\Phi_0$ are not included; while 
the initial data for those solutions in which $\Phi$ takes the value $\Phi_0$ 
$n$-times are represented by $n$ different points. The latter can happen if 
$\Phi$ is not a monotonic function of the proper time $t$ (as we saw such a 
non-monotonic behaviour on Fig.~\ref{fig:BB} in subsection \ref{sub-4.3}, 
and an oscillatory behaviour on Fig.~\ref{fig:SBN1-Phi} in subsection 
\ref{sub-4.1}). Thus, apart from these potential multiple representations of 
initial states, this set of initial data could be considered as the $\Phi=
\Phi_0$ surface of the constraint hypersurface of the momentum phase space 
of the EccH system. We will call them `phase diagrams'.

\subsubsection{The structure of the set of the initial conditions with $\Phi_0
=\sqrt{6/\kappa}$}
\label{sub-4.4.2}

First, for the sake of simplicity, let us consider the case $\Phi_0=\sqrt{6/
\kappa}$. (Thus, with this choice, we \emph{a priori} excluded those 
solutions from our considerations for which the Higgs field remain smaller 
than $\sqrt{6/\kappa}$. Such are the solutions with Small Bang singularities, 
in which $\Phi$ takes the value $\sqrt{6/\kappa}$ at the \emph{singularity} 
rather than at regular points (see subsection \ref{sub-2.2}); and also the 
solutions with a Milne type singularity for the parameters $\phi_0\leq1$ 
(see subsections \ref{sub-2.3} and \ref{sub-2.4}). In subsection \ref{sub-4.5}
we consider three more complicated cases, when $\Phi_0<\sqrt{6/\kappa}$.) 
From the investigations of the regular asymptotic solutions in subsection 
\ref{sub-2.4} we know that only those points of the $(S_0,S_1)$-plane can 
represent initial values that are on or `above' the straight line $S_1=
\frac{2}{3}\chi_cS_0$; or, on or `below' the straight line $S_1=-\frac{2}{3}
\chi_cS_0$ (see the second inequality in (\ref{eq:reg})). In the states 
corresponding to the points of these lines, the canonical momentum $\Pi$ is 
vanishing. For the sake of simplicity, we discuss only the case $S_1\geq
\frac{2}{3}\chi_cS_0$. Hence, for each of the cases $k=0,\pm1$, the set of 
the initial conditions for the solutions with $S_1>0$ consists of two copies 
of the piece $S_1\geq\frac{2}{3}\chi_cS_0$ of the $(S_0,S_1)$-plane such that 
these two pieces are identified just along the line $S_1=\frac{2}{3}\chi_c
S_0$. The two copies correspond to the two signs $s=\pm1$ of the canonical 
momentum $\Pi$ at the $\Phi=\Phi_0$ hypersurface (see equation 
(\ref{eq:2.pi})). Therefore, the set of all the initial conditions for 
$\Phi^2_0=6/\kappa$ and given $k$ is the disjoint union of this set and the 
analogous one in which $S_1<0$. At this point it could perhaps be worth 
stressing that the actual parameters $S_0$ and $S_1$ for a given solution 
$(S(t),\Phi(t))$ (e.g. with a Big Bang type singularity) are read-off from 
the restriction of the solution to a neighbourhood of the $\Phi=\Phi_0$ 
hypersurface \emph{as a regular solution} (see subsection \ref{sub-2.4}), 
and these should not be confused with the expansion coefficients of the 
singular solution itself near the singularity. E.g. the $S_0$ for the latter 
would be zero by assumption.

\subsubsection{The `phase diagram' of the solutions with $\Phi_0=\sqrt{6/
\kappa}$}
\label{sub-4.4.3}

Using the 1-parameter family of singular solutions with the (power series) 
Big Bang singularity for $k=0,\pm1$ or with the Milne type singularity (in 
which case $k=-1$ and $\phi_0>1$), we can determine the value of $(S,\dot S,
\dot\Phi)$ at the instant when $\Phi=\sqrt{6/\kappa}$ at first time. From 
these we can read off $S_0$, $S_1$ and $\phi_1$ (see subsection 
\ref{sub-2.4}). However, by equation (\ref{eq:2.14a}), these quantities are 
not independent, and this equation can be considered as a quadratic algebraic 
equation for $(S_1/S_0)$. Comparing its solution with the value $(S_1/S_0)$ 
obtained directly from the numerical solution we can determine the sign $s$. 
In this way we obtain a 1-parameter family of points in the $(S_0,S_1)$-plane 
corresponding to the solutions with (power series) Big Bang singularity in 
all the three cases $k=0,\pm1$; and another one corresponding to solutions 
with the Milne type singularity. In all the solutions with the Big Bang 
singularity we found $s=-1$ (or, more precisely, $s$ is typically between 
$-0.99$ and $-1.01$, although the numerical uncertainties near the limit 
line is greater); but in the Milne case $s$ can be either of $\pm1$. 

\begin{figure}[ht]
\begin{center}
\includegraphics[width=0.45\textwidth]{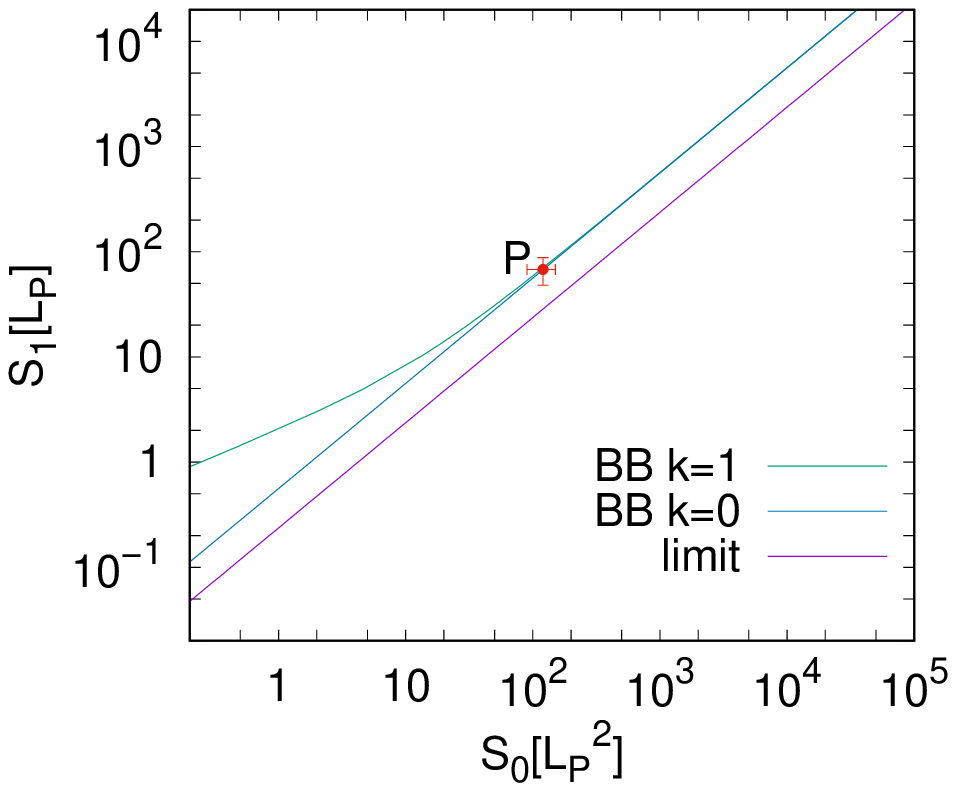}
\includegraphics[width=0.45\textwidth]{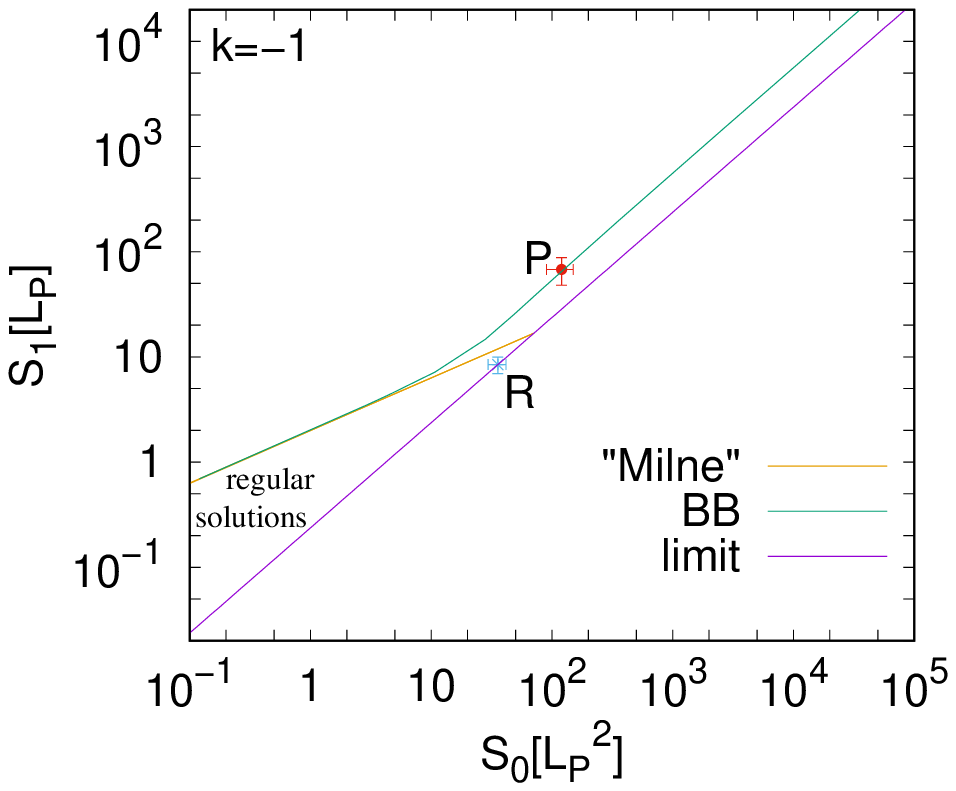}
\caption{\label{fig:Regular}
The $(S_0,S_1)$-plane of the initial conditions for the solutions with a Big 
Bang (BB) or Milne type singularity with $k=0,1$ (left figure) and $k=-1$ 
(right figure) at the moment when $\Phi=\sqrt{6/\kappa}$ at first time. All 
solutions correspond to points above the (purple) limiting straight line. 
The data for solutions with a (power series) Big Bang singularity are only 
on the $s=-1$ copy of the $(S_0,S_1)$-plane. For the initial data 
corresponding to solutions with a Milne type singularity $k=-1$ but $s=\pm1$. 
The 1-parameter family of these initial data on the $s=1$ and $s=-1$ copies 
of the $(S_0,S_1)$-plane coincide, and the one on the $s=1$ copy is continued 
in the other on the $s=-1$ copy. The points between the two Milne lines on the 
$s=-1$ and $s=1$ sheets, including the limit line (and the point $R$ on it), 
are initial conditions for regular solutions. Possibly apart from points of a 
subset of measure zero, all the other points appear to correspond to singular 
solutions in which the singularity cannot be reached asymptotically by a 
power series. Such an initial state is represented by the point $P$ on the 
$s=1$ sheet.}
\end{center}
\end{figure}

The data corresponding to the solutions with Big Bang for the $k=1$ and 
$k=0$ cases can be drawn on the same diagram; and those with Big Bang with 
$k=-1$ and with Milne type singularity on another one (see 
Fig.~\ref{fig:Regular}). 

In case of solutions with a power series Big Bang type singularity (see 
subsection \ref{sub-3.2}), as the parameter $\phi^4_{-1/2}$ of the asymptotic 
power series solution increases towards its critical value $\kappa/12\lambda$, 
the corresponding values of $(S_0,S_1)$ on the `phase diagram' tend to 
infinity. This regime of the parameter, $\phi^4_{-1/2}<\kappa/12\lambda$, 
corresponds to $k=1$ (see Fig.~\ref{fig:Regular}, left panel). The critical 
value of this parameter, $\phi^4_{-1/2}=\kappa/12\lambda$, corresponds to 
another 1-parameter family of solutions. For these solutions $k=0$, and in 
the $(S_0,S_1)$-plane they form a straight line that is the asymptote of the 
family of the $k=1$ data sets. This line appears to be parallel with the 
limit straight line $S_1=\frac{2}{3}\chi_cS_0$. In fact, in subsection 
\ref{sub-3.2} we saw that, near the singularity, the solution with $k=0$ 
depends on the parameter in a trivial way: The parameter is a simple overall 
scale factor. The numerical calculations show that this feature remains 
characteristic even on a much larger scale. 

After passing the critical value $\kappa/12\lambda$ of the parameter 
$\phi^4_{-1/2}$ (and hence when already $k=-1$) the corresponding initial 
values draw the same (at least numerically indistinguishable) line on the 
$(S_0,S_1)$-plane (see Fig.~\ref{fig:Regular}, right panel). However, in 
the domain $\phi^4_{-1/2}>\kappa/12\lambda$ of the parameter the values 
$(S_0,S_1)$ are \emph{decreasing} with increasing $\phi^4_{-1/2}$. As we 
already noted, the initial data for any solution with power series Big Bang 
singularity are on the $s=-1$ copy of the $(S_0,S_1)$-plane. 

Next, let us consider the initial data for solutions with a Milne type 
singularity (see subsection \ref{sub-2.5}). As the parameter $\phi_0$ of the 
asymptotic solutions tends from below to a special value $\phi_0^0$ (whose 
value, up to numerical uncertainties, is between $1.9$ and $2.1$), the 
corresponding curve is on the $s=1$ copy of the $(S_0,S_1)$-plane, and it 
approaches the limiting line $S_1=\frac{2}{3}\chi_cS_0$. Increasing $\phi_0$ 
further and passing the special value $\phi_0^0$, the curve continues on the 
$s=-1$ copy of the $(S_0,S_1)$-plane (see Fig.~\ref{fig:Regular}, right 
panel). The numerical calculations show that, on the $s=-1$ copy, the Milne 
line is an asymptote of the Big Bang line when $S_0\rightarrow0$. 

Finally, note that for $k=-1$ there are regular solutions, i.e. which do not 
have a singularity neither in the future nor in the past direction. The 
corresponding points on the `phase diagram' are between the two Milne lines 
(i.e. for which $s=-1$ and $s=1$), including the limit straight line $S_1=
\frac{2}{3}\chi_cS_0$. In particular, the point $R$ corresponds to the special 
regular solution with initial data $(S_c^2,2\chi_cS^2_c/3)$, discussed at the 
end of subsection \ref{sub-2.4} (see Fig.~\ref{fig:Regular}, right panel).

\subsubsection{An evidence for the existence of non-power series 
asymptotic solutions}
\label{sub-4.4.4}

\begin{figure}[ht]
\begin{center}
\includegraphics[width=0.45\textwidth]{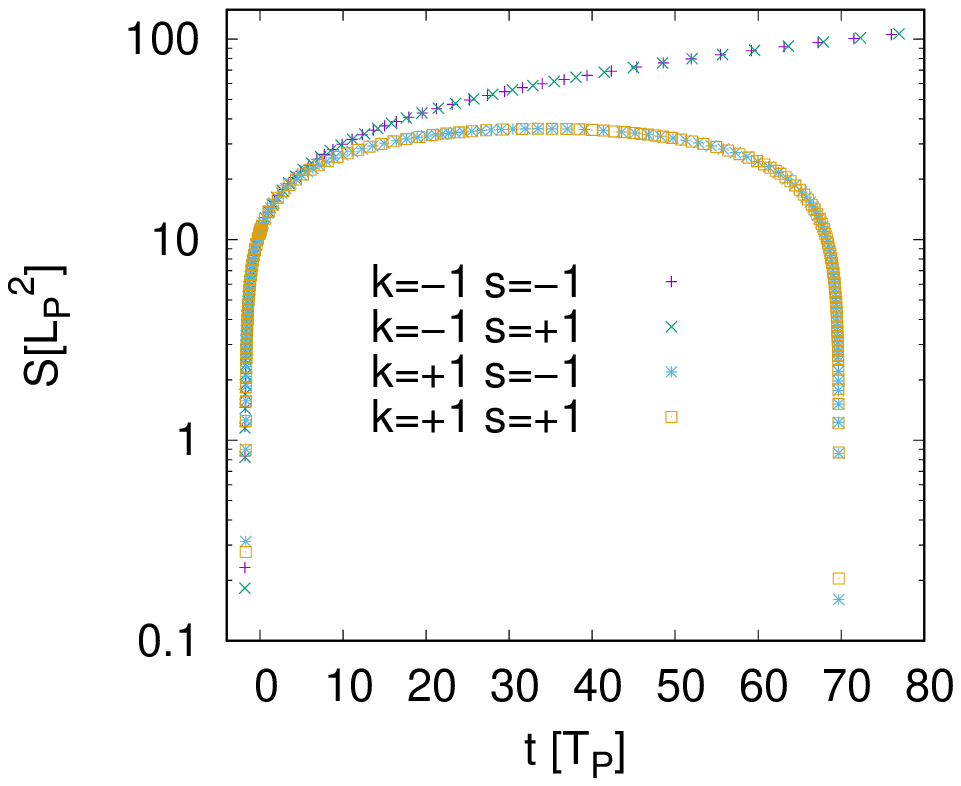}
\includegraphics[width=0.45\textwidth]{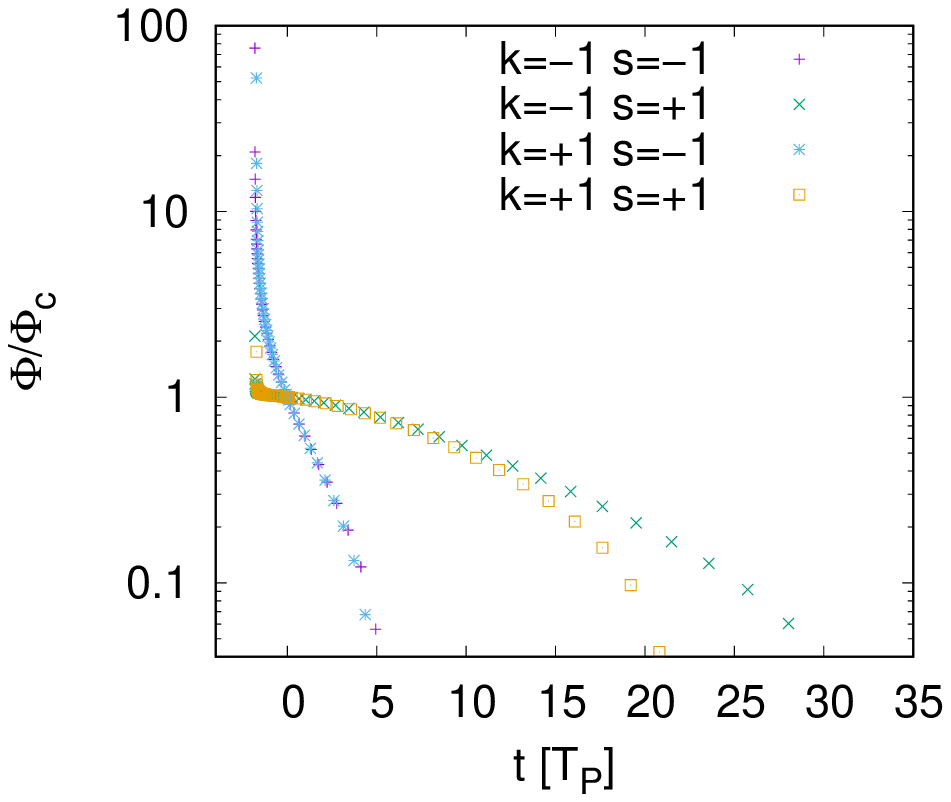}
\caption{\label{fig:s01203s10680}
The scale function and the Higgs field, as a function of time (both for $t<0$ 
and $t>0$), in the numerical solution developed from the initial condition 
$(S_0,S_1)=(120.2L_P^2,68.0L_P)$ (denoted on Fig.~\ref{fig:Regular}, by $P$) 
at $t=0$ with $k=\pm 1$ and $s=\pm1$. The plus, the cross, the star and the 
box denote, respectively, the $(k,s)=(-1,-1)$, the $(-1,1)$, the $(1,-1)$ and 
the $(1,1)$ cases. $\Phi_c$ denotes $\sqrt{6/\kappa}$.}
\end{center}
\end{figure}

To clarify the nature of the solutions with initial data not lying on the 
distinguished lines above, let us consider the point $(S_0,S_1)=(120.2 L_P^2,
68.0 L_P)$ of the $(S_0,S_1)$-plane, denoted by $P$ on Fig.~\ref{fig:Regular}. 
This corresponds both to the values $\phi^2_{-1/2} = 1.94$ and 
$\phi^2_{-1/2}=2.2$ of the parameter of the solution with a (power series) 
Big Bang singularity (with $k=1$ and $k=-1$, respectively) \emph{provided} 
$s=-1$. (On the `phase diagram' this point is separated enough from the 
Milne line, because the numerical error for determining the $(S_0,S_1)$ 
values for a Big Bang solution is definitely less than the parameter 
distance of the point above from the Milne line.) However, with this initial 
data at $t=0$ we can solve the evolution equations, both in the past ($t<0$) 
and future ($t>0$) directions, \emph{even for $s=1$, too}. The results for 
all the four possibilities $k=\pm1$, $s=\pm1$ are shown in 
Fig.~\ref{fig:s01203s10680}. 

The left panel of Fig.~\ref{fig:s01203s10680} shows the time evolution of 
the scale function $S$. This function appears to be independent of the sign 
of $s$ (which is true for all other initial points from the Big Bang line 
that we studied). On the other hand, in the right panel, we can see that 
while for $s=-1$ and $k=\pm1$ the Higgs field $\Phi$ smoothly diverges as $t$ 
decreases from zero, but for $s=1$ and $k=\pm1$ we get an abrupt increase 
of $\Phi$ just below $t=0$ (though the two solutions for $k=1$ and $k=-1$ 
are different). These solutions start with $S=0$ (at some $t<0$) and a value 
of $\Phi$ which is finite or infinite (although the numerical calculations 
indicate that it is probably infinite). If this starting value of $\Phi$ 
were finite, then the solutions would be of Milne type. However, this could 
be possible only if some Milne type singularity could not be reached by a 
power series type asymptotic solution, because their initial values $(S_0,
S_1)$ are not on the Milne line. Similarly, if for $s=1$ these solutions 
started with $S=0$ and $\Phi=\infty$, then these solutions could not have 
power series Big Bang singularities either, because for them $s=-1$ would 
hold. Therefore, we obtained an example for a solution with either a Milne 
or (probably) Big Bang type singularity in which the singularity cannot be 
reached by a power series (`non-analytic Big Bang').

\subsubsection{The nature of solutions with a generic initial data with 
$\Phi_0=\sqrt{6/\kappa}$}
\label{sub-4.4.5}

Finally, let us discuss what kind of solutions do we obtain from a generic 
point (above the limiting line) of the $(S_0,S_1)$-plane. It turned out that 
the singular or non-singular nature of the solutions is independent of the 
sign of $s$. In particular, if $k=1$, then all the solutions appear to start 
and end in a singularity. (We inspected solutions with sixteen randomly 
chosen initial data sets on the $(S_0,S_1)$-plane.) In addition to the line 
indicated on the left panel of Fig.~\ref{fig:Regular} (representing a 
1-parameter family of solutions with power series type Big Bang singularity), 
there may be more such lines. They would correspond to multiple 
representations of solutions with power series type Big Bang singularity 
(see the remark at the end of subsection \ref{sub-4.4.1}). The rest of the 
two dimensional allowed subset of the two copies of the $(S_0,S_1)$-plane 
appears to correspond to solutions which are non-analytic around their 
singularity. 

If $k=0$, then all the solutions appear to start ($S_1>0$) or end ($S_1<0$) 
in a singularity. The line corresponding to solutions with power series type 
Big Bang singularity is a straight line (see the left panel of 
Fig.~\ref{fig:Regular}). The rest of the two dimensional allowed subset of 
the $(S_0,S_1)$-plane seems to correspond to solutions which are non-analytic 
around their singularity. 

If $k=-1$, then either a solution is regular everywhere, or it starts or 
ends in a singularity. The union of the `wedges' between the Milne-line and 
the limiting line on the $s=1$ and $s=-1$ copies of the $(S_0,S_1)$-plane 
contains initial data for the regular solutions (see the right panel of 
Fig.~\ref{fig:Regular}). In particular, the special solution with the data 
$(S_0,S_1)=(S^2_c,\frac{2}{3}\chi_cS^2_c)$ (denoted by $R$ on 
Fig.~\ref{fig:Regular}, right panel) is regular. Indeed, the fact that the 
boundary of the set of initial data for the regular solutions is just the 
Milne line could be expected: The Milne singularity is a singularity only 
of the \emph{solution}, but not of the spacetime geometry (see subsections 
\ref{sub-2.3} and \ref{sub-2.5}). All the points outside this open set appear 
to represent initial data for singular solutions; and, apart from the lines 
corresponding to solutions with a power series Big Bang or Milne type 
singularity, these are non-analytic around the singularities.

\subsection{Solutions with initial data on hypersurfaces with $\Phi<
\sqrt{6/\kappa}$}
\label{sub-4.5}

\subsubsection{The structure of the set of the initial conditions with $\Phi_0
<\sqrt{6/\kappa}$}
\label{sub-4.5.1}

Choosing $\Phi_0$ in the definition of the initial hypersurface (of 
subsection \ref{sub-4.4.1}) to be \emph{less} than $\sqrt{6/\kappa}$, e.g. to 
be $\frac{1}{2}\sqrt{6/\kappa}$, $\frac{3}{5}\sqrt{6/\kappa}$ or $\frac{4}{5}
\sqrt{6/\kappa}$ (as in the following numerical calculations), we include 
initial data for solutions with (power series) Small Bang singularities in our 
set of initial conditions, too. This makes the `phase diagram' of the 
solutions slightly more complicated, but the analysis of subsections 
\ref{sub-4.4.2}--\ref{sub-4.4.4} can be repeated. 

In fact, for given $\Phi=\Phi_0$ the value of $S$, $\dot S$ and $\dot\Phi$ 
are not independent, because they must satisfy the constraint in 
(\ref{eq:1.1}) (together with the expression (\ref{eq:1.3}) for the energy 
density). This yields the second order algebraic equation 

\begin{equation}
\dot\Phi^2+2\Phi_0\bigl(\frac{\dot S}{S}\bigr)\dot\Phi+\Phi_0^2\bigl(\mu^2+
\frac{\Lambda}{3}+\frac{1}{2}\lambda\Phi^2_0\bigr)-(\frac{6}{\kappa}-
\Phi^2_0)\Bigl(\bigl(\frac{\dot S}{S}\bigr)^2+\frac{k}{S^2}-\frac{\Lambda}{3}
\Bigr)=0 \label{eq:4.1}
\end{equation}
for $\dot\Phi$. To have a solution of this equation, its discriminant 
must be non-negative, 

\begin{equation}
D:=\frac{6}{\kappa}\bigl(\frac{\dot S}{S}\bigr)^2-\Phi^2_0\bigl(\mu^2+
\frac{\Lambda}{3}+\frac{1}{2}\lambda\Phi^2_0\bigr)+(\frac{6}{\kappa}-
\Phi^2_0)\Bigl(\frac{k}{S^2}-\frac{\Lambda}{3}\Bigr) \ge 0,  \label{eq:4.2}
\end{equation}
in which case the solution of (\ref{eq:4.1}) is 

\begin{equation}
\dot\Phi=-\Phi_0(\frac{\dot S}{S})\pm\sqrt{D}. \label{eq:4.3}
\end{equation}
The sign in front of $\sqrt{D}$ will be denoted by $s$. 

For a given $\Phi^2_0$ condition (\ref{eq:4.2}) could be a non-trivial 
constraint, but for a different $\Phi^2_0$ it could be satisfied identically. 
In particular, for $\Phi^2_0=6/\kappa$ condition (\ref{eq:4.2}) reduces to 
the second inequality in (\ref{eq:reg}), in which case it is independent of 
the discrete parameter $k$; and the boundary $D=0$ of the initial conditions 
for solutions, i.e. the limit line, is a straight line on the $(S,
\dot S)$-plane. However, for $\Phi^2_0\not=6/\kappa$ this condition depends 
on the value of $k$, and the boundary $D=0$ is given by 

\begin{equation}
\dot S^2=-k(1-\frac{\kappa}{6}\Phi^2_0)+\bigl(\frac{\kappa}{6}\mu^2\Phi^2_0
+\frac{\kappa\lambda}{12}\Phi^4_0+\frac{\Lambda}{3}\bigr)S^2. \label{eq:4.2a}
\end{equation}
Here we do not give the exhaustive discussion of the quite diverse 
possibilities, simply we illustrate some of them by a few examples. The 
detailed analysis is elementary and straightforward. In particular, for $k=0$ 
(\ref{eq:4.2a}) gives two straight lines on the $(S,\dot S)$-plane through 
the origin, and the coefficient of $S^2$ on the right hand side is positive 
precisely when either 

\begin{equation}
\lambda\Phi^2_0<-\mu^2-\sqrt{\mu^4-4\Lambda\lambda/\kappa} 
\hskip 20pt {\rm or}  \hskip 20pt 
-\mu^2+\sqrt{\mu^4-4\Lambda\lambda/\kappa}<\lambda\Phi^2_0. \label{eq:4.4}
\end{equation}
Thus, if $\lambda\Phi_0^2$ is chosen to satisfy (\ref{eq:4.4}), then only 
those points of the $(S,\dot S)$-plane with $\dot S\geq0$ can represent 
initial data that are on or `above' the limit line. Any of our present choice 
for $\Phi^2_0$ will be much bigger than $(-\mu^2+\sqrt{\mu^4-4\Lambda\lambda/
\kappa})/\lambda$, and hence (\ref{eq:4.2}) is a non-trivial constraint with 
a non-trivial limit (straight) line (see e.g. Fig.~\ref{fig:intersection05}, 
right panel). Therefore, the set of the initial data with $\dot S\geq0$ 
consists of two copies of the part of the $(S,\dot S)$-plane `above' the 
limit line, labelled by the two signs $s=\pm1$, which copies are identified 
along the limit line. 

For $k=1$ and $\Phi^2_0$ making the coefficient of $S^2$ in (\ref{eq:4.2a}) 
positive, (\ref{eq:4.2a}) gives a nontrivial limit curve for any $S^2$ only 
for $\Phi^2_0>6/\kappa$; but for $\Phi^2_0<6/\kappa$ (as in our case), the 
limit curve starts from the point $(S_{\bullet},0)$ both in the $\dot S>0$ and 
$\dot S<0$ half planes, where 

\begin{equation*}
S^2_{\bullet}:=2\frac{6-\kappa\Phi^2_0}{\kappa\lambda\Phi^4_0+2\kappa\mu^2
\Phi^2_0+4\Lambda}.
\end{equation*}
In this case the $(S,\dot S)$-plane does not split into the two disjoint 
pieces $\dot S>0$ and $\dot S<0$, they join together along the line between 
the points $(0,0)$ and $(S_{\bullet},0)$. The set of all the initial data 
consists of two such copies, labelled by the sign $s=\pm1$, which are 
identified along their limit curves (see e.g. Fig.~\ref{fig:intersection05}, 
middle panel). 

If $k=-1$, then an analogous analysis yields that, for $\Phi^2_0<6/\kappa$ but 
making the coefficient of $S^2$ in (\ref{eq:4.2a}) positive, the limit curve 
has two disconnected branches (one in the $\dot S>0$ and the other in the 
$\dot S<0$ half-plane), exists for any $S>0$, and they start at $(0,\pm\sqrt{1-
\kappa\Phi^2_0/6})$. The set of all the initial data with $\dot S>0$ consists 
of two copies of the points on or `above' the limit curve, labelled by the 
sign $s=\pm1$, which are identified along their limit curves (see e.g. 
Fig.~\ref{fig:intersection05}, left panel). The set of all the initial data 
is the disjoint union of this set and the one with $\dot S<0$.

\subsubsection{The `phase diagram' of the solutions with $\Phi_0<\sqrt{6/
\kappa}$}
\label{sub-4.5.2}

\begin{figure}[ht]
\begin{center}
\includegraphics[width=0.32\textwidth]{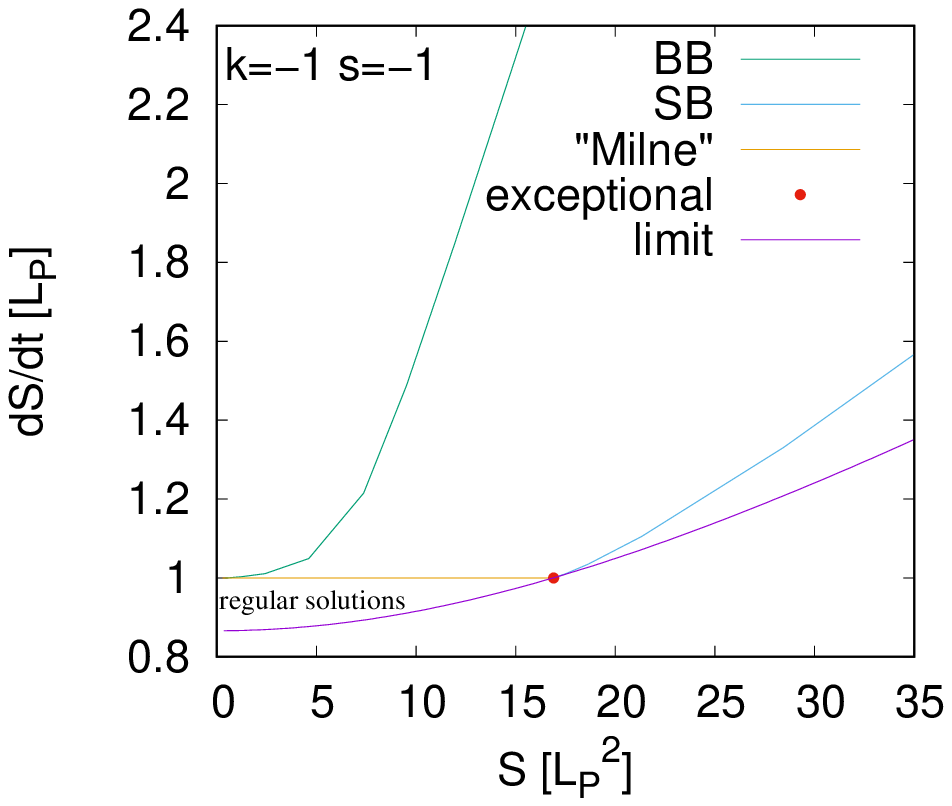}
\includegraphics[width=0.32\textwidth]{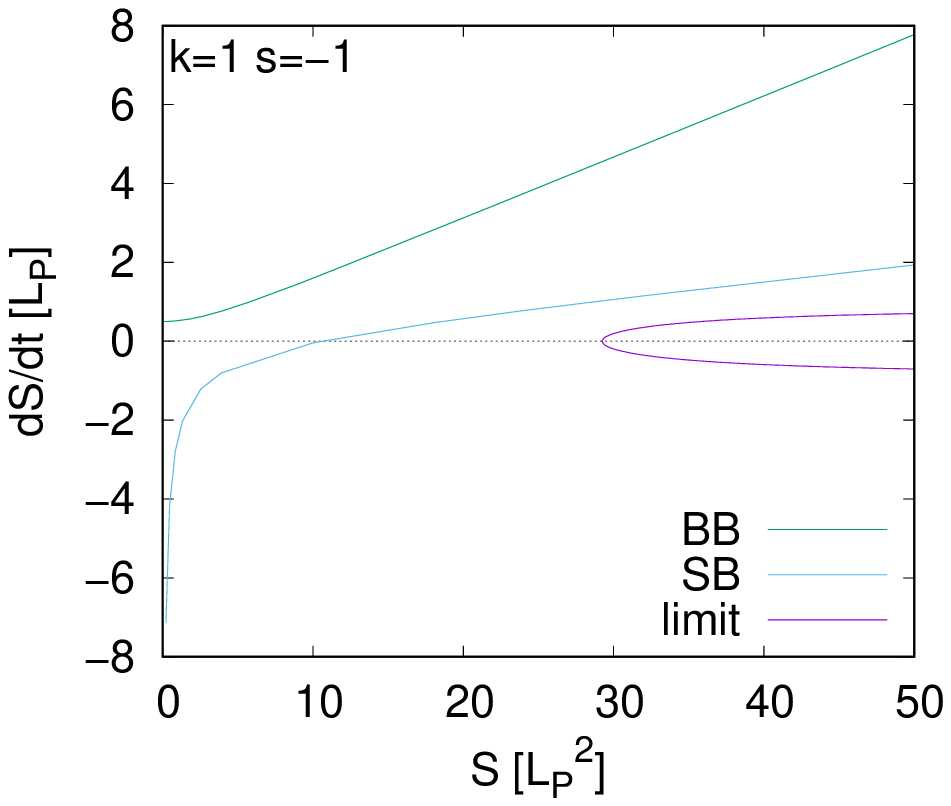}
\includegraphics[width=0.32\textwidth]{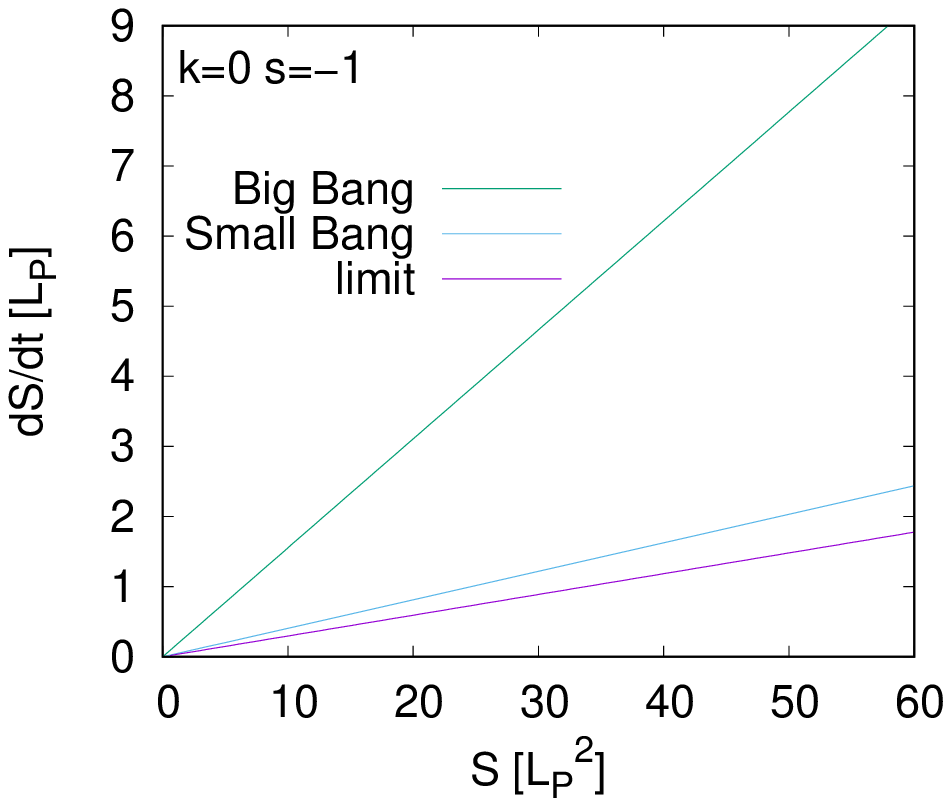}
\caption{\label{fig:intersection05}
The `phase diagram' for $\Phi_0=\frac{1}{2}\sqrt{6/\kappa}$ and $k=-1,1$ and 
$0$, respectively. BB is the Big Bang, and SB is the Small Bang line.}
\end{center}
\end{figure}

\begin{figure}[ht]
\begin{center}
\includegraphics[width=0.32\textwidth]{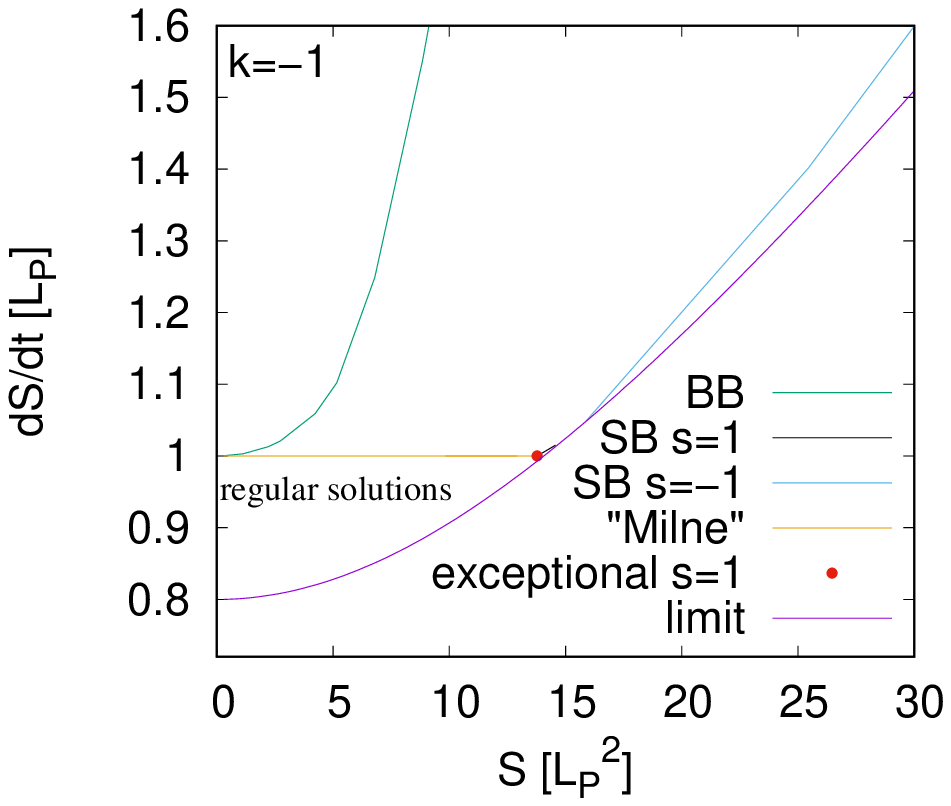}
\includegraphics[width=0.32\textwidth]{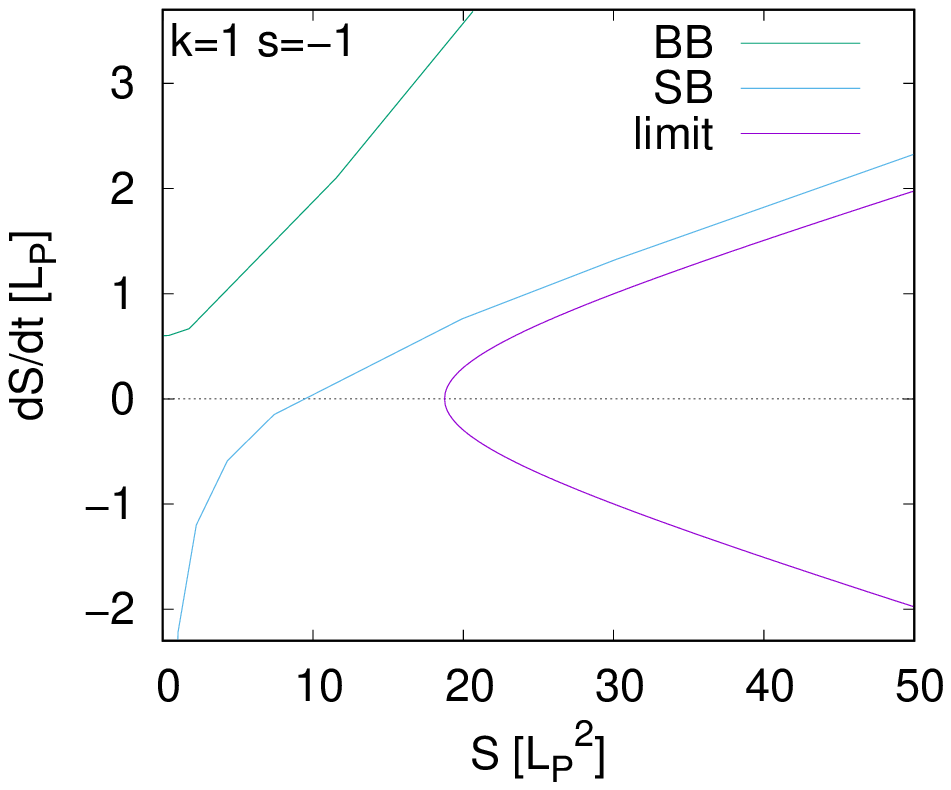}
\includegraphics[width=0.32\textwidth]{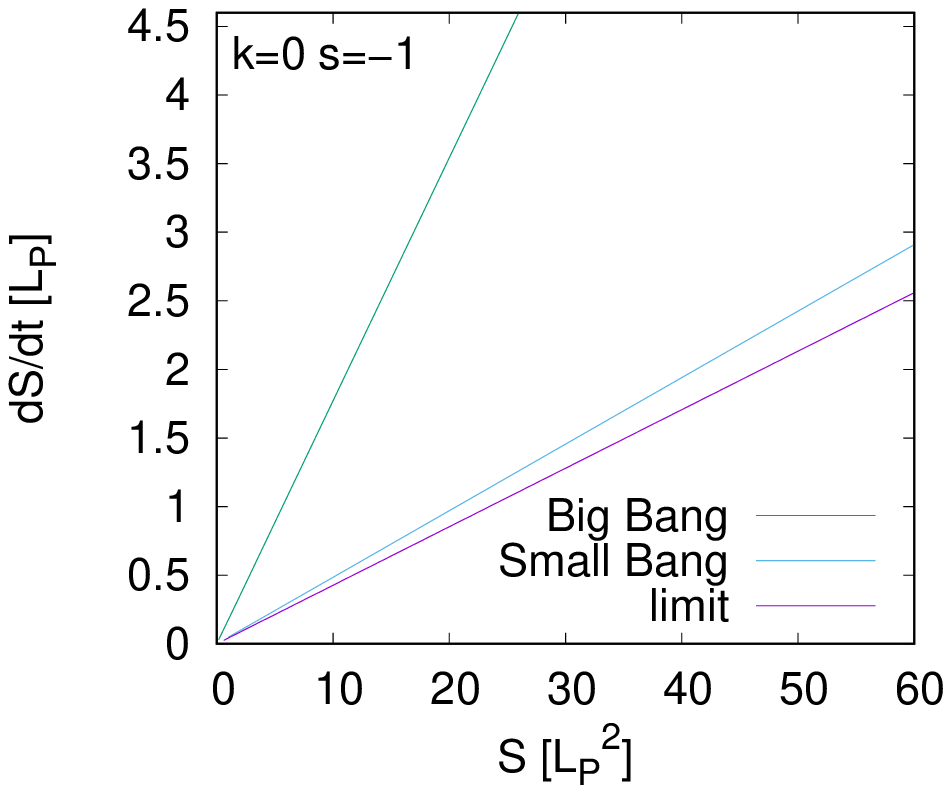}
\caption{\label{fig:intersection06}
The `phase diagram' for $\Phi_0=\frac{3}{5}\sqrt{6/\kappa}$ and $k=-1,1$ and 
$0$, respectively. The Big Bang (BB) line is on the $s=-1$ copy of the 
$(S,\dot S)$-plane. SB is the Small Bang line.}
\end{center}
\end{figure}

\begin{figure}[ht]
\begin{center}
\includegraphics[width=0.32\textwidth]{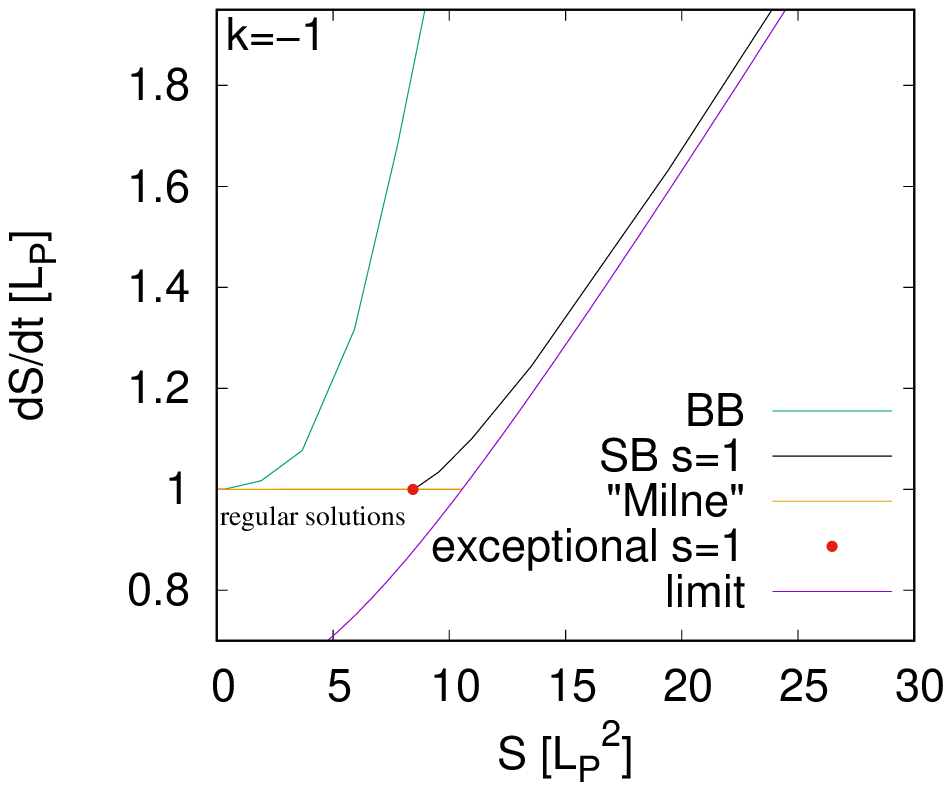}
\includegraphics[width=0.32\textwidth]{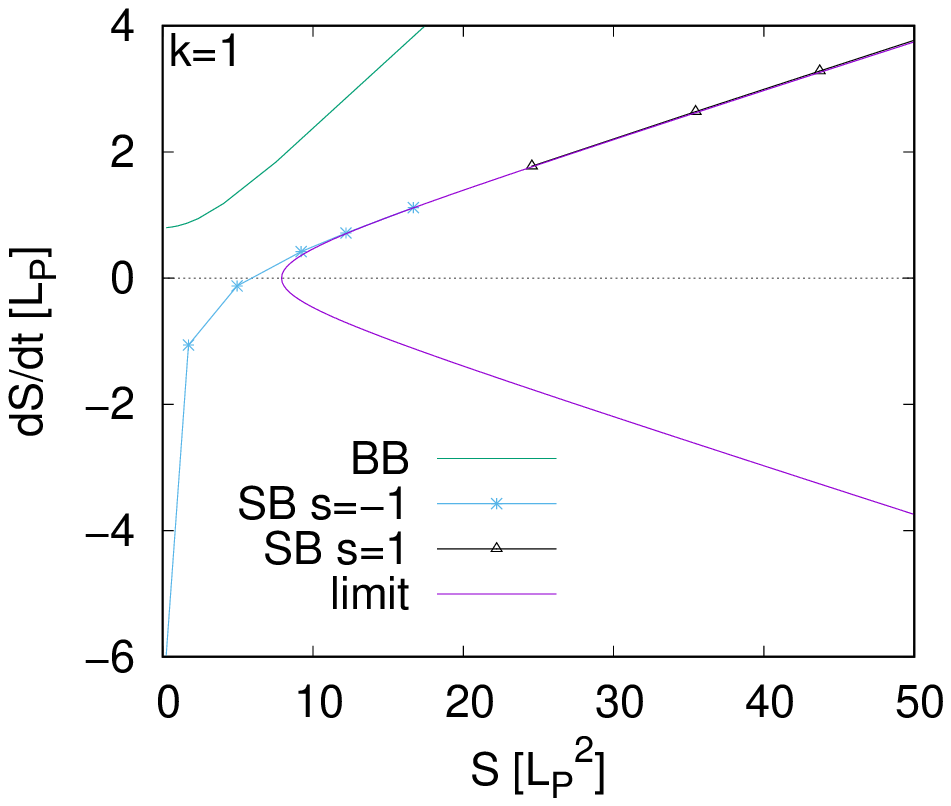}
\includegraphics[width=0.32\textwidth]{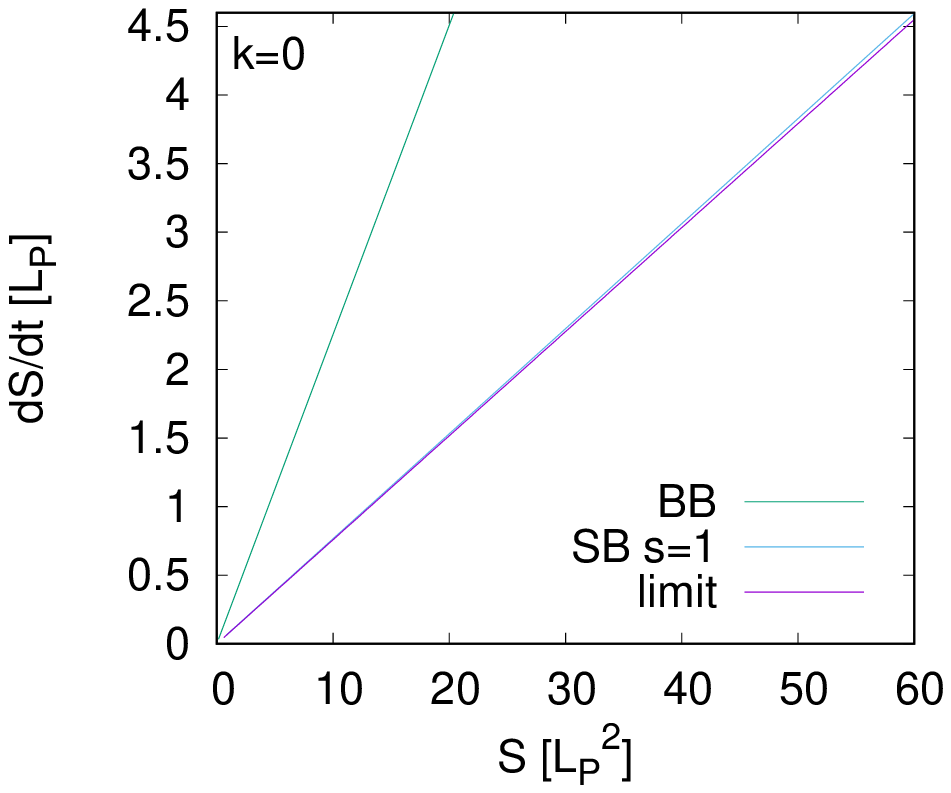}
\caption{\label{fig:intersection08}
The `phase diagram' for $\Phi_0=\frac{4}{5}\sqrt{6/\kappa}$ and $k=-1,1$ and 
$0$, respectively. The Big Bang (BB) line is on the $s=-1$ copy of the 
$(S,\dot S)$-plane. SB is the Small Bang line. }
\end{center}
\end{figure}

Similarly to subsection \ref{sub-4.4.3}, starting from the singularity of a 
singular solution, we determine the initial data on the spacelike 
hypersurface specified by the condition that $\Phi$ takes the value $\Phi_0=
\frac{1}{2}\sqrt{6/\kappa}$, or $\frac{3}{5}\sqrt{6/\kappa}$, or $\frac{4}{5}
\sqrt{6/\kappa}$ for the first time. Now, to parametrize these solutions, we 
still use the parameters appearing in their asymptotic form near the 
singularity (and discussed in subsections \ref{sub-2.2}, \ref{sub-2.5} and 
\ref{sub-3.2}). Comparing the solution (\ref{eq:4.3}) with the value 
$(\dot S/S)$ obtained directly from the numerical solution we can determine 
the sign $s$. In this way, in any of the three choices for $\Phi_0$, we 
obtain 1-parameter families of points in the $(S,\dot S)$-plane corresponding 
to solutions with the (power series) Big Bang (BB) and Small Bang (SB) 
singularities in all the three cases $k=0,\pm1$; and a 1-parameter family of 
points corresponding to solutions with the Milne type singularity (with 
$k=-1$). (The exceptional solution of subsection \ref{sub-2.3} corresponds to 
a single point on the Milne line.) The results are shown, in the $\Phi_0=
\frac{1}{2}\sqrt{6/\kappa}$, $\frac{3}{5}\sqrt{6/\kappa}$ and $\frac{4}{5}
\sqrt{6/\kappa}$ cases, respectively, by Fig.~\ref{fig:intersection05}, 
Fig.~\ref{fig:intersection06} and Fig.~\ref{fig:intersection08}. 

We saw that for $\Phi_0=\sqrt{6/\kappa}$ the Big Bang lines in the $k=1$ and 
$k=-1$ cases on the $(S_0,S_1)$-plane (numerically) coincided, for $k=0$ this 
was a straight line through the origin, and in all these cases we obtained 
$s=-1$. For any solutions with a power series Big Bang singularity $s=-1$ 
still holds even for $\Phi_0<\sqrt{6/\kappa}$, and, for $k=0$, this is still 
a straight line, but these lines for $k=-1$ and $k=1$ do not coincide any 
more. In addition to this, the only essential, qualitative difference between 
the $\Phi_0=\sqrt{6/\kappa}$ and the present cases is that now solutions with 
Small Bang singularities are present, and the parameter domain for the 
solutions with a Milne type singularity is enlarged. Thus, we discuss only 
these in detail.

Solutions with a Milne singularity exist only for $k=-1$, and the main 
qualitative properties of the Milne line in the present $\Phi_0<\sqrt{6/
\kappa}$ cases appear to be the same that we saw when $\Phi_0$ was $\sqrt{6/
\kappa}$ (see the left panel of Fig.~\ref{fig:intersection05}, 
Fig.~\ref{fig:intersection06} and Fig.~\ref{fig:intersection08}). The only 
new phenomenon is that the exceptional solution (with asymptotics discussed 
in subsection \ref{sub-2.3} and corresponding to the parameter $\phi_0=1$), 
a special member of the Milne family, appears: For $\Phi_0=\frac{4}{5}
\sqrt{6/\kappa}$ the corresponding point in the `phase diagram' is on the 
$s=1$ copy of the $(S,\dot S)$-plane (see Fig.~\ref{fig:intersection08}), 
and decreasing $\Phi_0$ to tend to $\frac{1}{2}\sqrt{6/\kappa}$ this point 
seems to tend to the limit line (see Fig.~\ref{fig:intersection05}). For 
smaller $\Phi_0$ we expect this point to be already on the $s=-1$ copy of 
the $(S,\dot S)$-plane.

However, the initial data for solutions with a (power series) Small Bang 
singularity have much more complicated structure. First, for $k=-1$ the 
Small Bang line tends asymptotically, in the $S_1\rightarrow0$ limit, to the 
initial data for the exceptional solution, but, in contrast to the Big Bang 
line, it is not confined into the $s=-1$ copy of the $(S,\dot S)$-plane: If 
$\Phi_0=\frac{1}{2}\sqrt{6/\kappa}$, then the Small Bang line is in the 
$s=-1$ copy, but for greater $\Phi_0$ it starts (asymptotically) from the 
point for the exceptional solution on the $s=1$ leaf, crosses the limit 
line, and then continues on the $s=-1$ leaf (see the left panel of 
Fig.~\ref{fig:intersection05}, Fig.~\ref{fig:intersection06} and 
Fig.~\ref{fig:intersection08}).

If $k=1$, then, as we saw in subsection \ref{sub-4.5.2}, the `phase diagram' 
is connected, it does not split into the disjoint pieces $\dot S>0$ and 
$\dot S<0$. In fact, for any of the given values for $\Phi_0<\sqrt{6/\kappa}$, 
a piece of the Small Bang line is in the $\dot S<0$ domain, another is in the 
$\dot S>0$ domain, and there is an initial state in which $\dot S=0$. This 
latter is the data for the solution in which the universe starts to recollapse 
just when the Higgs field takes the value $\Phi_0$. The piece of the Small Bang 
line in the $\dot S<0$ domain is on the $s=-1$ copy of the $(S,\dot S)$-plane. 
Increasing $\Phi_0$ to tend to $\sqrt{6/\kappa}$, the Small Bang line is 
getting to be closer and closer to the limit line. After crossing the limit 
line, the Small Bang line continues on the $s=1$ leaf (see the middle panels 
of Fig.~\ref{fig:intersection05}, Fig.~\ref{fig:intersection06} and 
Fig.~\ref{fig:intersection08}). 

As we saw in subsection \ref{sub-2.2}, for $k=0$ the parameter $S_1$ is only 
an overall scale factor of the asymptotic solution, and the Higgs field is 
independent of $S_1$. The numerical calculations demonstrate this behaviour 
on a much larger scale, independently of the value of $\Phi_0$: On the $(S,
\dot S)$-plane the line corresponding to these solutions is a straight line 
through the origin (see the right panels). On the other hand, the slope of 
the Small Bang straight line depends on the value of $\Phi_0$: Increasing 
$\Phi_0$ to tend to to $\sqrt{6/\kappa}$, the Small Bang line is getting to 
be closer and closer the limit line. For $\Phi_0$ less than a special value 
(which is approximately $0.78\,\sqrt{6/\kappa}$) the Small Bang line is still 
on the $s=-1$ leaf. but increasing $\Phi_0$ further it is already on the 
$s=1$ leaf (see the right panels of Fig.~\ref{fig:intersection05}, 
Fig.~\ref{fig:intersection06} and Fig.~\ref{fig:intersection08}). 

Apart from the domains for the regular solutions and the lines corresponding 
to solutions with (Big Bang, Small Bang or Milne type) singularities that 
can be reached by power series, the points correspond to initial values for 
singular solutions that are not analytic near their singularity.

\section{Conclusions and summary}
\label{sec:5}

We investigated the Einstein--conformally coupled Higgs field (EccH) system 
in the presence of Friedman--Robertson--Walker symmetries both analytically 
(near the initial singularities) and numerically. We determined analytically 
all the asymptotic, power series solutions up to fourth order near the 
singularities. We found three 1-parameter families of solutions. In the first 
both the Higgs field and certain scalar polynomial curvature invariants 
diverge (Big Bang), in the second the Higgs field remain bounded but certain 
scalar polynomial curvature invariants diverge (Small Bang), and in the third 
both the Higgs field and the curvature invariants remain bounded. In fact, 
while the first two are genuine physical spacetime singularities; the third 
is only a Milne type singularity, and the spacetime can be extended to a 
bigger one through this. The existence of these singular solutions 
demonstrates that the symmetry breaking instantaneous vacuum states of the 
Higgs sector are not only kinematical possibilities, but that, as it was 
claimed in \cite{Sz16}, they \emph{do} emerge non-trivially during the 
dynamics of the system. 

We determined these solutions numerically, starting from the sub-Planck 
scale to the era of the weak interactions, as well. We found that the 
asymptotic, power series solutions above give surprisingly good approximation 
even on this scale. Also, we investigated numerically the generic properties 
of the set of the initial data of the EccH system. The solutions with the 
Big Bang, Small Bang and Milne type singularities above turned out to form 
only a \emph{subset of measure zero} in the set of all the initial conditions. 
The complement of these is the union of the set of initial conditions for the 
regular solutions (with $k=-1$), and that for \emph{singular solutions that 
cannot be expanded in power series near the singularities}.

\bigskip
\noindent
Gy. Wolf was supported by the Hungarian OTKA fund K109462.



\begin{thebibliography}{99.}%

\bibitem{HE} S. W. Hawking, G. F. R. Ellis, {\it The Large Scale Structure of 
             Spacetime}, Cambridge University Press, Cambridge 1973

\bibitem{AL73} E. S. Abers, B. W. Lee, Gauge theories, Phys. Rep. {\bf 9} 
               1--141 (1973)

\bibitem{Sz16} L. B. Szabados, Gravity, as a classical regulator for the 
               Higgs field, and the genesis of rest masses and electric charge, 
               arXiv: 1603.06997v3 

               L. B. Szabados, On gravity's role in the genesis of rest masses 
               of classical fields, Gen. Relat. Grav. (2018) {\bf 50} 34, 
               DOI:10.1007/s10714-018-2340-1, arXiv: 1802.04401 

\bibitem{ES} G. F. R. Ellis, B. G. Schmidt, Singular spacetimes, Gen. Rel. 
             Grav. {\bf 8} 915-953 (1977)

\bibitem{num-rec} W. H. Press, S. A. Teukolsky, W. T. Vetterling,
               B. P. Flannery, {\it Numerical Recipies in Fortran}
                Cambridge University Press, 1992
             
\end{thebibliography}
\end{document}